\documentclass[10pt,twocolumn]{article}

\usepackage[margin=0.78in]{geometry}
\usepackage{float}
\usepackage{graphicx}
\graphicspath{{images/}}
\usepackage{epstopdf}
\usepackage{amsmath}
\usepackage{placeins}
\usepackage{bm} 
\usepackage{amsthm}

\usepackage{mathtools}
\usepackage{amssymb}
\usepackage[english]{babel}
\usepackage{graphicx}
\usepackage{sidecap}
\usepackage{caption}
\usepackage{multicol}
\usepackage{multirow}
\usepackage{tabularx}
\usepackage{blindtext, subfig}
\usepackage{dblfloatfix} 
\usepackage[utf8]{inputenc}
\usepackage[table]{xcolor}
\usepackage[square, numbers, comma, sort&compress]{natbib}
\usepackage{hyperref}
\usepackage[switch]{lineno}
\definecolor{mygrayfonce}{gray}{0.4}
\definecolor{myorange}{RGB}{255, 223, 191}

\title{Photon-noise: Is a single-pixel camera better than point scanning? A signal-to-noise ratio analysis for Hadamard and Cosine positive modulation.}

\usepackage{authblk}
\author[1]{Camille Scott\'e}
\author[1]{Fr\'ed\'eric Galland}
\author[1]{Herv\'e Rigneault}

\affil[1]{Aix Marseille University, CNRS, Centrale Marseille, Institut Fresnel, Marseille, France}

\date{}

\begin{document}

\maketitle

\textbf{Abstract-} In a single-pixel camera, an unknown object is sequentially illuminated by intensity patterns. The total reflected or transmitted intensity is summed in a
single-pixel detector from which the object is computationally reconstructed. In the situation where the measurements are limited by photon-noise, it is questionable whether a single-pixel camera  performs better or worse than simply scanning the object with a focused intensity spot - a modality known as \textit{point raster scanning} and employed in many laser scanning systems. Here, we solve this general question and report that positive intensity modulation based on Hadamard or Cosine patterns does not necessarily improve the single-to-noise ratio (SNR) of single-pixel cameras as compared to point raster scanning. Instead, we show that the SNR is only improved on object pixels at least $k$ times brighter than the object mean signal $\bar{x}$, where $k$ is a constant that depends on the modulation scheme. This fundamental property is demonstrated theoretically and numerically. It is also experimentally confirmed in the spatial domain - for widefield fluorescence imaging - and in the spectral domain - for spontaneous Raman spectral measurements. Finally, we provide user-oriented guidelines that help decide when and how multiplexing under photon-noise should be used instead of point raster scanning. \\

\begin{figure} [h]
\centering
\includegraphics[width=\linewidth]{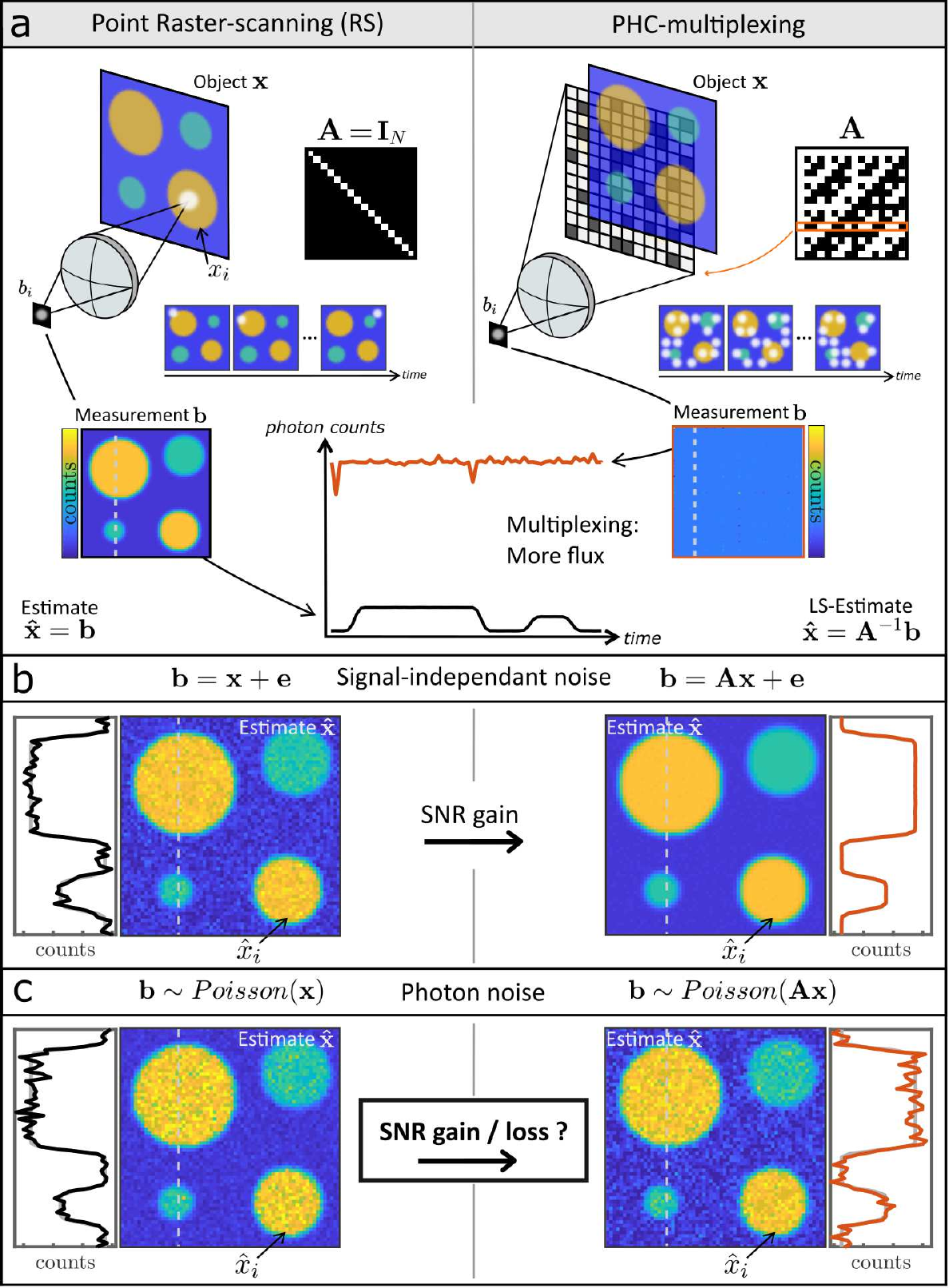} 
\caption {\textbf{a}, Schematic representation of Point raster-scanning and Positive Hadamard or Cosine-multiplexing in the absence of noise. At fixed irradiance and exposure time, PHC-multiplexing detects more photons. $\mathbf{x}$: intensity
object; $\mathbf{\hat{x}}$: estimation of $\mathbf{x}$ from the measurements $\mathbf{b}$; $\mathbf{A}$: multiplexing matrix (real and positive, invertible); $\mathbf{I}_N$: identity matrix; LS: least-square. 
\textbf{b}, Example of estimated object in the presence of signal-independent noise, modelled as additive white Gaussian noise. \textbf{c}, Example of estimated object, for photon noise.}
\label{fig:ConceptFig}
\end{figure}

Over the last decade, single-pixel cameras have received increasing attention in fields as diverse as microscopy \cite{Radwell2014a,Lochocki2016a}, spectroscopy \cite{Berto2017}, photoacoustic imaging \cite{Huynh2019} or cytometry \cite{Ota2018}. Single-pixel cameras, combined with various computational techniques, offer the promise of considerably faster and cheaper optical systems \cite{Davenport2012, Edgar2019}. A single-pixel camera typically relies on some form of \textit{multiplexing}. Unlike \textit{point raster-scanning (RS)} - where an object is probed point-by-point - in multiplexing the signal from different parts of an object is combined into a single-pixel detector (Fig.~\ref{fig:ConceptFig} a). The object is thus seen through a sequence of time-varying patterns and the detected signal must be demultiplexed to retrieve the original information. Here, we consider intensity modulation multiplexing (measurements are incoherent sums of intensities), achieved via the widely used Hadamard or Cosine-based positive patterns. This type of multiplexing is referred to as \textit{PHC: Positive-Hadamard and Cosine multiplexing}. \\

\noindent  A major asset of PHC-multiplexing is known as the \textit{Multiplexing advantage} \cite{Fellgett1967}: It is an improvement in signal-to-noise ratio (SNR) brought by multiplexing over RS (Fig.~\ref{fig:ConceptFig} b), when the measurement noise comes from the detector electronics (additive signal-independent noise). Then, multiplexing via Hadamard or Cosine based-patterns leads to the detection of consequently more signal than RS, thereby comparatively reducing the amount of noise and dramatically improving the SNR (Fig.~\ref{fig:ConceptFig} a-b). 
This property of PHC-multiplexing has been known since the 1960s \cite{Jacquinot1964a, Roland1967, 1385, Fellgett1967, Treffers1977, Harwit1979, Decker1971a, DeVerse2000}; but it is with the concomitant advent of spatial-light-modulators, efficient computational imaging techniques, and high-speed and low-noise detectors that multiplexing with single-pixel detectors became extremely popular \cite{Schechner2003, Sun2013, Sun2016a, Radwell2014, Pian2017, Zhang2017a, Scotte2018, Moshtaghpour2018, Xiang2011a, Garbacik2018, Wijesinghe2019, Toninelli2020, Scotte2019, Scotte2020, ScotteThesis2020, Zhang2015, Studer2012}. \\

\noindent One consequence of using high-performance single-pixel detectors is that their noise may be so low that the main source of noise in the system now arises from the photon-counting process itself (Fig.~\ref{fig:ConceptFig} c). Yet, in this photon-noise regime, the multiplexing advantage does not hold any more \cite{Minami1987, Larson1974, Schumann2002, Wuttig2005, Streeter2009a}: PHC-multiplexing does not ensure a SNR improvement over RS (Fig.~\ref{fig:ConceptFig} c). 
This effect was partially studied in a few dated works \cite{Roland1967, Larson1974, Hirschfeld1976, Kahn2002, Bialkowski1998, Fuhrmann2004, Shin2013},
which only considered average SNR values, and therefore do not enable one to conclude if, yes or no, and when and how, PHC-multiplexing is beneficial over RS in terms of SNR. Despite the attention that multiplexing has received in recent years, this fundamental question has remained largely unaddressed or ignored. Therefore, in a context of increasing use of computational-imaging techniques based on multiplexing such as compressive sensing \cite{Candes2006a, Duarte2008, Marcia2011} or ghost imaging \cite{Bromberg2008}, and with progress in detectors technology that tends to make measurements more and more likely to be limited by photon noise
only \cite{Hadfield2020}, we believe it is necessary to clarify under which circumstances PHC-multiplexing brings a SNR advantage over RS, for photon-noise limited data (Fig.~\ref{fig:ConceptFig} c). \\

\noindent In this paper, we theoretically, numerically, and experimentally compare raster-scanning and PHC-multiplexing, in terms of signal-to-noise ratio, when the noise only arises from the photon-counting process.
We show that, even when PHC-multiplexing leads to the detection of consequently more photons than RS, it does not necessarily improve the SNR, and can even degrade it significantly. More precisely, we show that, on a given object, PHC-multiplexing only improves the SNR of object parts brighter than a certain threshold value that depends on the multiplexing implementation strategy and on the sample average signal. This allows us to draw user-oriented guidelines that help decide  when and how PHC-multiplexing should be used instead of RS.
The results presented in this paper are supported by a supplementary methods and a detailed auxiliary paper that provides theoretical proofs \cite{Scotte2022}.

\subsection*{Model and Assumptions}
\label{sec: Model_and_assumptions}

Although not limited to a specific dimensionality or experimental system, for the sake of clarity and without loss of generality, we base the narrative on the specific example of a simple incoherent 2-D imaging system such as in Fig.~\ref{fig:ConceptFig} a, made of (i) an intensity object $\mathbf{x}$, (ii) an optical lens for signal collection, (iii) a single-pixel detector. 
Fig.~\ref{fig:ConceptFig} b illustrates the well-know multiplexing advantage mentioned in the introduction: when the noise is additive and independent of the signal, both positive-Hadamard and positive-Cosine multiplexing substantially improve the SNR as compared to RS, by a factor proportional to $\sqrt{N}$, with $N$ the number of pixels \cite{Jacquinot1964a, Fellgett1967, Harwit1979, Decker1971a, DeVerse2000}. 
Fig.~\ref{fig:ConceptFig} c illustrates the case where the noise arises from the photon-counting process. There, the intensity of an object $\mathbf{x}$ modulated through a positive matrix $\mathbf{A} \in \mathbb{R}_{+}^{N \times N}$, leads to measurements $\mathbf{b}$: 
\begin{equation} 
\mathbf{b} \sim Poisson(\mathbf{Ax}) 
\label{eqn:completemodel}
\end{equation}
where $\mathbf{x}$ and $\mathbf{b}$ are assumed to be real and positive quantities. The object $\mathbf{x}$ contains the intensities $x_i$ from every pixel $i$ ($\mathbf{x} = [x_1,…,x_N]^T$), and the measurement $\mathbf{b}$ contains the observed photon counts $b_i$ ($\mathbf{b} = [b_1,…,b_N]^T$). Each measured number of photons $b_i$ is a random variable whose probability law is a Poisson distribution of mean $\langle b_i \rangle = [\mathbf{Ax}]_i$. 
$\mathbf{A}$ is the multiplexing matrix that contains the positive modulation patterns and is assumed to be invertible. In RS, $\mathbf{A}$ is the identity matrix $\mathbf{I}_N$ (each measurement $b_i$ collects signal from a single object pixel $i$). 
We further assume that (i) the measurements as statistically independent, (ii) the number of measurements is equal to the number of probed object pixels $N$, (iii) that the system optical resolution is smaller than the finest object structures. 
Note that, unless otherwise stated, all results hold for any object dimensionality (1-D, 2-D, etc.) - as long as the variables can be rearranged in the form of equation~\eqref{eqn:completemodel} - and for any experimental system - as long as it complies with the linear model of equation~\eqref{eqn:completemodel} and its assumptions. \\

\noindent \textbf{Comparison metrics: }
Since the measurements $\mathbf{b}$ are noisy, one cannot perfectly access the ground-truth object $\mathbf{x}$ but can only estimate it. This estimate, denoted $\hat{\mathbf{x}}$ , is directly equal to the measurements for raster-scanning, and to their demodulation for multiplexing. In both cases, it differs from $\mathbf{x}$ by some error $\delta\hat{\mathbf{x}} = \hat{\mathbf{x}} - \mathbf{x}$. The aim of this work is to determine which of RS or PHC-multiplexing lead to the smallest error.
This is assessed with the mean-square error (MSE) and signal-to-noise ratio (SNR). Both inform on how precise and accurate is the estimate $\hat{\mathbf{x}}$  on each object pixel $i$: 
\begin{equation}
MSE(\hat{x}_i) = \langle(\hat{x}_i - x_i)^2 \rangle 
\text{ ; }
SNR(\hat{x}_i) = \frac{x_i}{\sqrt{MSE(\hat{x}_i)}} 
\label{eqn:def_mse} 
\end{equation}
The potential SNR improvement or degradation brought by PHC-multiplexing over RS can then be quantified with the following ratio:
\begin{equation} \label{eqn:def_G}
G_i = \frac{SNR(\hat{x}_i)_{multiplex}}{SNR(\hat{x}_i)_{RS}} =
\sqrt{\frac{MSE(\hat{x}_i)_{RS}}{MSE(\hat{x}_i)_{multiplex}}}
\end{equation}
If $G_i>1$, multiplexing improves the SNR on pixel $i$ as compared to RS, and conversely. Since SNR and MSE are directly related, and to bypass the additional dependence on the object ground-truth, in the following we only give results in terms of MSE. \\ 

\noindent \textbf{Multiplexing matrices}: The SNR depends on the multiplexing matrix. Here, we focus on
positive-Hadamard multiplexing and on positive-Cosine modulation with the discrete-cosine transform (DCT). These two widely used classes of multiplexing are generally implemented via matrices with coefficients comprised between 0 and 1.  \\

\begin{figure} [hbt!]
\centering
\includegraphics[width=\linewidth]{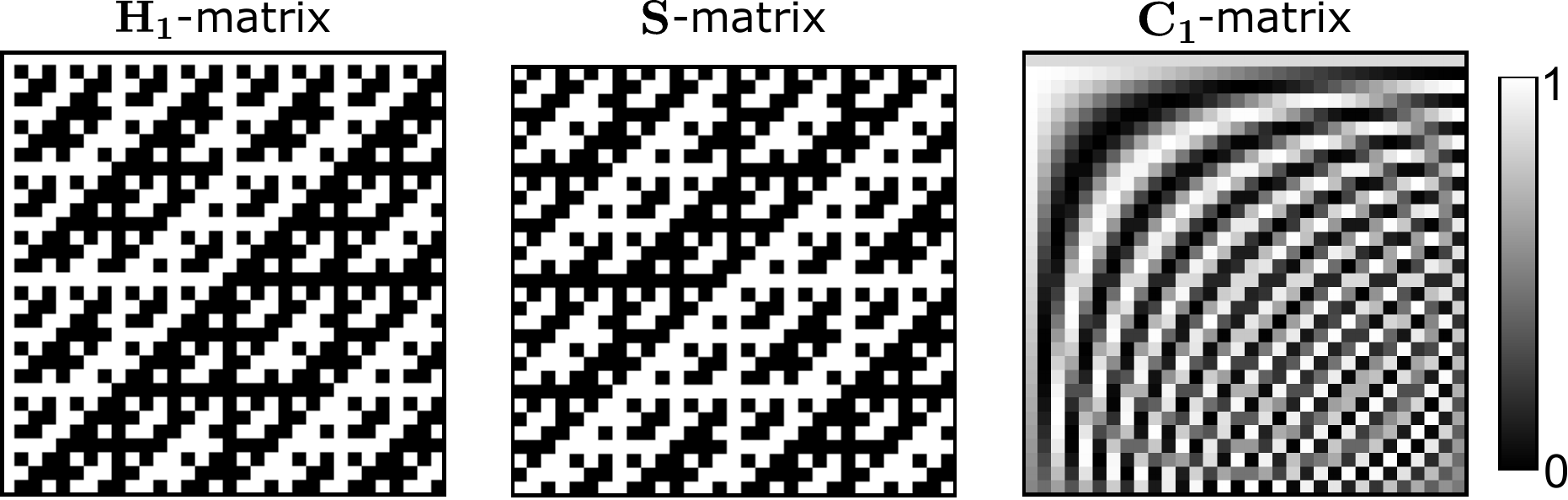} 
\caption {Considered multiplexing matrices. $\mathbf{H_1}$  and $\mathbf{S}$ are associated with positive-Hadamard multiplexing and $\mathbf{C_1}$ with positive-Cosine multiplexing }
\label{fig:3matrices}
\end{figure}

\noindent \textit{Positive-Hadamard multiplexing} (Fig.~\ref{fig:3matrices}) is often implemented by modulating or blocking parts of the light with simple absorptive patterns \cite{Harwit1979} or with light modulator devices \cite{DeVerse2000, Studer2012, Scotte2020}, as schematically depicted in Fig.~\ref{fig:ConceptFig}a. The associated multiplexing matrix is binary, and can for instance be: (i) the matrix $\mathbf{H_1}$ (Hadamard matrix with $-1$ elements replaced with $0$):
\begin{equation}
\mathbf{H_1} = \frac{1}{2}(\mathbf{H}+\mathbf{J})
\label{eqn:def_H1}
\end{equation}
where  $\mathbf{H}$ is the Hadamard matrix $\mathbf{J}$ is the constant matrix of ones; or 
(ii) the matrix $\mathbf{S}$ (e.g. Hadamard matrix without first row and column, with $-1$ elements replaced with $+1$, and $+1$ elements $0$), defined via \cite{Harwit1979}:
\begin{equation} 
\left\{
\begin{array}{ll}
\mathbf{S}^T\mathbf{S} = \mathbf{S}\mathbf{S}^T = \frac{N+1}{4}(\mathbf{I} + \mathbf{J}) \\
\mathbf{J} \mathbf{S} = \mathbf{S}\mathbf{J} = \frac{N+1}{2}\mathbf{J}
\end{array}
\right.
\label{eqn:Smatrixpty}
\end{equation}
For both matrices, about half of the $N$ coefficients of each row are ones, and half are zeros (Fig.~\ref{fig:3matrices}). More details are provided in \cite{Scotte2022}. 
\\

\noindent \textit{Positive-Cosine multiplexing} (Fig.~\ref{fig:3matrices}) can be implemented in different ways (e.g. \cite{Zhang2015, Zhang2017, Meng2020, Futia2011, Scotte2019}). 
In this text, we exclusively focus on positive-cosine intensity modulation based on the matrix $\mathbf{C_1}$ of equation~\eqref{eqn:C1DCT_def}. It is based on the DCT \cite{Strang1999}, and defined such that the coefficients of $\mathbf{C_1}$ are comprised between 0 and 1:
\begin{equation}
\mathbf{C_1} = \frac{1}{2} (\mathbf{DCT} + \mathbf{J})
\label{eqn:C1DCT_def}
\end{equation}
where $\mathbf{DCT}$ is the discrete-Cosine transform matrix with coefficients comprised between $-1$ and $+1$ (see \cite{Scotte2022} for other positive-cosine modulation types).  

\begin{figure} [hbt!]
\centering
\includegraphics[width=\linewidth]{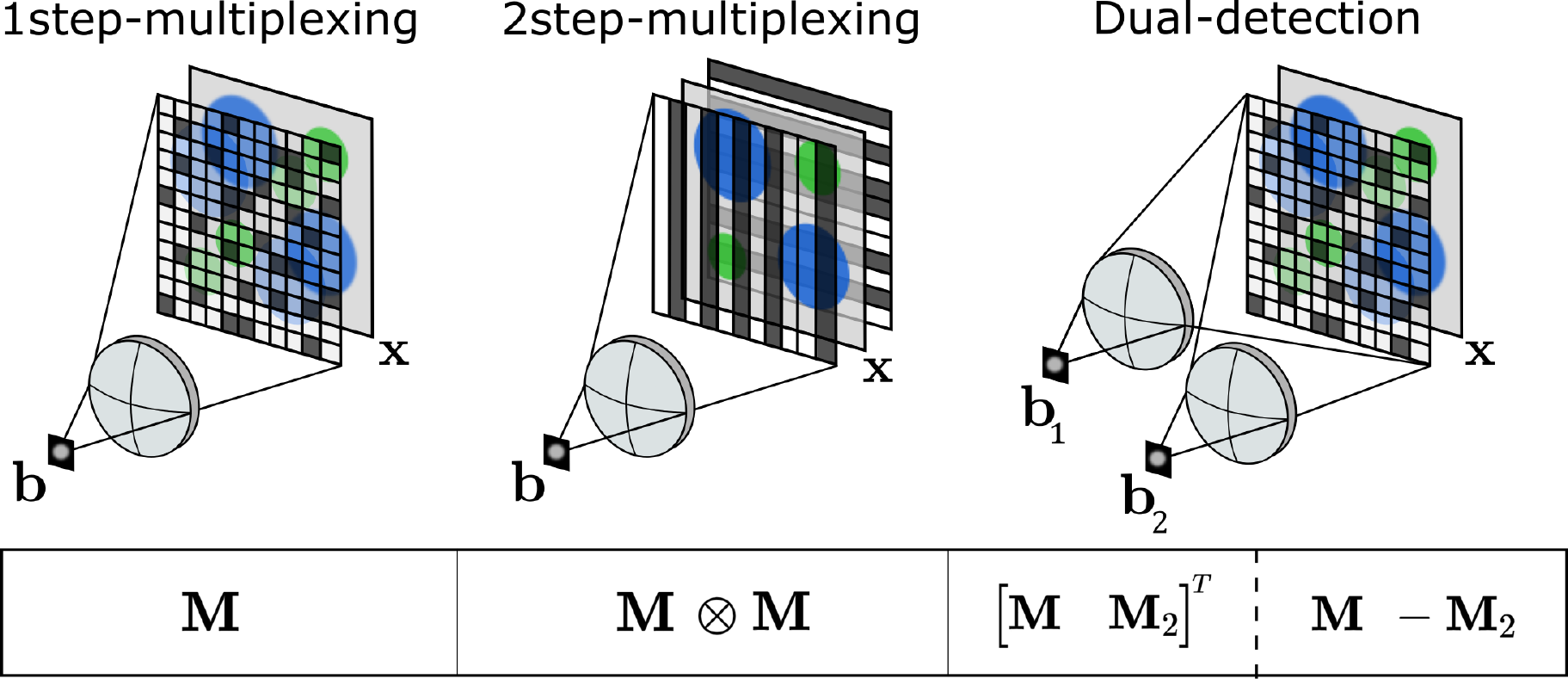} 
\caption {Schematic in 2-D of the considered positive-multiplexing schemes, with their associated equivalent multiplexing matrix. In dual-detection, the measurements can be combined into a single vector (left) or subtracted (right). $\mathbf{M_2}$: complementary matrix of $\mathbf{M}$.} 
\label{fig:Modalities_MainTxt}
\end{figure}

\noindent \textbf{Multiplexing schemes}: The SNR may also depend on the chosen single-pixel multiplexing scheme. Here, we consider three configurations, illustrated in 2-D in Fig.~\ref{fig:Modalities_MainTxt} and detailed in \cite{Scotte2022} section 3. \\

\noindent In \textit{One-step multiplexing}, an object is probed with a series of patterns which have the same dimensionality as the object. Each measurement $b_i$ is then the sum of the point-wise product between the object and a pattern encoded in the $i^{th}$ row of the multiplexing matrix (Fig~\ref{fig:ConceptFig} and Fig.~\ref{fig:Modalities_MainTxt}).  \\

\noindent \textit{Two-step multiplexing} applies only in 2-D: A 2-D object can be multiplexed with two independent 1-D multiplexing stages that probe uncorrelated dimensions, such as the vertical and horizontal dimensions. The two sets of patterns derive from the rows and columns of two distinct multiplexing matrices, and the equivalent multiplexing matrix is their Kronecker product (Fig.~\ref{fig:Modalities_MainTxt}). \\

\noindent In \textit{Dual-detection}, the one-step multiplexing scheme is supplemented with an additional detector, such that the two detectors make complementary measurements $\mathbf{b_1}$ and $\mathbf{b_2}$ (the non-collected signal by the first detector is collected by the second detector). This is equivalent to associating the matrix $\mathbf{M}$ to the measurements $\mathbf{b_1}$ and the matrix $\mathbf{M_2}$ to the measurements $\mathbf{b_2}$, such that: $\mathbf{M} + \mathbf{M_2} = \mathbf{J}$ ($\mathbf{J}$ is the matrix of ones).
The measure can be reconstituted by combining the measurements from each detector into a single vector; or by subtracting them (Fig.~\ref{fig:Modalities_MainTxt}). The later approach, often found in the literature \cite{Enk2019, Zhang2015, Soldevila2016, Meng2020}, is referred to as \textit{Balanced detection}. \\

\noindent \textbf{Number of photons: } 
The SNR depends on the number of photons collected by RS and PHC-multiplexing. To begin with, we draw the comparison when the number of photons is \textit{not} constant. Rather, we give the advantage to multiplexing by comparing them at fixed exposure time and irradiance. On the example of Fig.~\ref{fig:ConceptFig}, this means each sample pixel is illuminated with the same light power: if in raster-scanning, each pixel of the object is illuminated with 1 mW during 1 ms, then in multiplexing, each pixel of the object will also see 1 mW of incident light during 1 ms. This results in a consequently higher measured number of photons for PHC-multiplexing (Fig.~\ref{fig:ConceptFig}a): for $N$ measurements, if raster-scanning leads to a total of $N\bar{x}$ photon counts, PHC-multiplexing leads to a total of about $\frac{N^2}{2}\bar{x}$ photon counts. \\ 

\noindent \textbf{Estimation method}:
The SNR depends on the estimator used to demultiplex the raw measurements. To begin with, we estimate $\hat{\mathbf{x}}$ with the least-square (LS) estimation, i.e. via $\mathbf{\hat{x}} = \mathbf{A}^{-1} \mathbf{b}$ (see Supplementary Methods). 


\subsection*{Results}
\label{sec: Calculs-Simulations}
In this context, we prove (see mathematical derivations in \cite{Scotte2022} section 4) that, for the three considered multiplexing schemes, and both positive-Hadamard multiplexing and positive-Cosine multiplexing; the MSE obtained with least-square estimation is approximately constant over the estimated object $\mathbf{\hat{x}}$, on most pixels $i$ and for a number of pixels $N \gg 1$. The MSE is proportional to the average signal contained in the object $\bar{x}$:
\begin{equation}
MSE_{PHC}(\hat{x}_i) \approx k \bar{x}
\label{eqn:MSE_kx}
\end{equation}
where $k$ is a positive constant that depends on the multiplexing scheme (Fig.~\ref{fig:Modalities_MainTxt}) and matrix (Fig.~\ref{fig:3matrices}). 
In opposite, in RS, the MSE equals the object itself: 
\begin{equation} 
MSE_{RS}(\hat{x}_i) = x_i
\label{eqn:MSE_RS}
\end{equation}
and the associated SNR scales with the square-root of the object intensity at each pixel $i$. Therefore, as compared to raster-scanning, PHC-multiplexing improves the SNR by a factor (equation~\ref{eqn:def_G}):
\begin{equation}
G_i = \sqrt{\frac{x_i}{k\bar{x}}}
\label{eqn:G_kx}
\end{equation}
Hence, PHC-multiplexing brings a SNR improvement over RS only on object pixels $i$ which intensity $x_i$ verify:
\begin{equation}
x_i \geq k \bar{x}
\label{eqn:ruleofthumb_kx}
\end{equation}
In other words, PHC-multiplexing only improves the SNR on pixels brighter that $k$ times the object mean signal $\bar{x}$, and degrades it on regions dimmer than this value. When averaged over all object pixels, the overall SNR is degraded by a factor $\sqrt{k}$, meaning that any SNR gain is compensated by a SNR loss on other pixels. Hence, although under our assumptions, PHC-multiplexing detects about $N/2$ times more photons than RS, it does not improve the SNR on every pixel of the estimated object. From this, it immediately appears that the choice between PHC-multiplexing and raster-scanning greatly depends on the value of $k$ (i.e. on the multiplexing scheme and matrix) and on the object structure (i.e. on how the object pixels are distributed as compared to the object average signal $\bar{x}$).\\ 

\noindent \textbf{Positive-Hadamard multiplexing: } 
Table.~\ref{tab:SumUp_HadaSmatCos} gives the theoretical values of $k$ for positive-Hadamard multiplexing, for the three multiplexing schemes of Fig.~\ref{fig:Modalities_MainTxt}. One-step multiplexing leads to a better theoretical MSE than two-step multiplexing, and the best MSE is achieved with a one-step multiplexing implemented with dual-detection. In general, the matrices $\mathbf{H_1}$ and $\mathbf{S}$ lead to the same MSE: One-step multiplexing leads to a MSE equals to twice the object average ($k=2$). Comparatively, two-step multiplexing degrades the MSE by a factor two ($k=4$); and dual-detection improves the MSE by a factor two ($k=1$). The differences between $\mathbf{H_1}$ and $\mathbf{S}$ lie (i) in the presence of few special pixels in the MSE for $\mathbf{H_1}$ (ii) in the dual-detection: indeed for dual-detection with $\mathbf{H_1}$, considering the full measurements or subtracting them lead to the same MSE, while with the $\mathbf{S}$-matrix, it is important \textit{not} to subtract the two measurements. The theoretical proofs for the MSE are derived in \cite{Scotte2022}, section 4.2. 
{\renewcommand{\arraystretch}{2}
\begin{table} [H]
\centering
\begin{tabularx}{\linewidth}{|p{0.7cm}|X|X|p{1.1cm}|p{1.1cm}|}
\hline
 &   One-step & Two-step  & \multicolumn{2}{|c|}{Dual-detection}  \\
\hline
\scalebox{0.9}{$\mathbf{A}$} &  \scalebox{0.9}{$\mathbf{M}$} & \scalebox{0.9}{$\mathbf{M} \otimes \mathbf{M}$}  & \scalebox{0.8}{$ \left[ \mathbf{M} \hspace{0.1cm} \mathbf{M_2} \right]^T$} & \scalebox{0.8}{$\mathbf{M}-\mathbf{M_2}$} \\ 
\hline \hline
\multicolumn{5}{|c|}{Positive-Hadamard multiplexing (\scalebox{0.9}{$\mathbf{M} = \mathbf{S}$)}} \\
\hline
\scalebox{0.9}{$MSE$}  & $2 \bar{x}$  $\scriptstyle (\forall i )$ & $4 \bar{x}$ $\scriptstyle (\forall i )$  & $ \bar{x}$  $\scriptstyle (\forall i )$ & $ 2\bar{x}$  $\scriptstyle (\forall i )$ \\
\hline
\multicolumn{5}{|c|}{Positive-Hadamard multiplexing (\scalebox{0.9}{$\mathbf{M} = \mathbf{H_1}$)}} \\
\hline
\scalebox{0.9}{$MSE$} &  $2 \bar{x}$  $\scriptstyle (\forall i \neq 1 )$  & $4 \bar{x}$  $\scriptstyle (\forall i \neq n_1)$ & \multicolumn{2}{|c|}{$ \bar{x}$   $\scriptstyle ( \forall i )$} \\
\hline
\end{tabularx}
\caption{Theoretical MSE for positive-Hadamard multiplexing. Results hold for LS-estimation and large number of pixels $N \gg 1$. The specific pixels $n_1$ and the theoretical proofs are given in \cite{Scotte2022} section 4.2. 
$\mathbf{A}$: equivalent multiplexing matrix. $\bar{x}$: object intensity average. }
\label{tab:SumUp_HadaSmatCos}
\end{table}
\begin{figure*} [hbt!] 
\centering
\includegraphics[width=\linewidth]{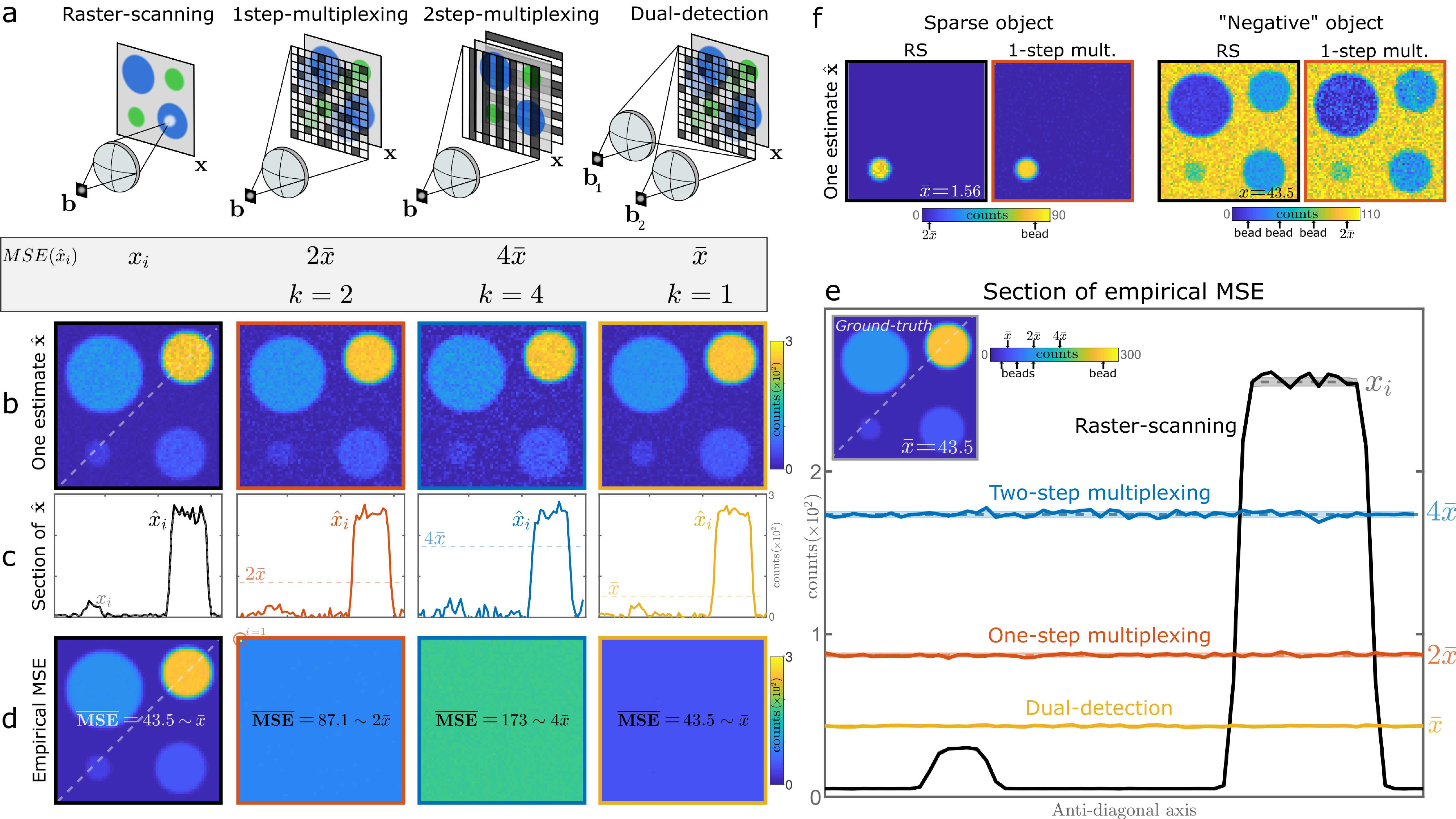} 
\caption{MSE for positive-Hadamard multiplexing and LS-estimation.
\textbf{a}, Schematic representation of a 2-D imaging system in point raster-scanning mode and in three multiplexing modes, with the associated theoretical MSE per object pixel $i$ and constant $k$ of equation~\eqref{eqn:MSE_kx}. Here, one-step multiplexing is performed with $\mathbf{H_1}$, two-step with $\mathbf{S} \otimes \mathbf{S}$, and dual detection with $\mathbf{H_1}-\mathbf{H_2}$ ($\mathbf{H_2}$ is the complementary of $\mathbf{H_1}$).
\textbf{b}, Example of an estimate $\hat{\mathbf{x}}$ obtained after one realisation of the data (one simulated measurement). For visualisation purposes, only positive estimated values are shown.
\textbf{c}, Section of $\hat{\mathbf{x}}$ along the anti-diagonal. The dashed-lines represent theoretical MSE values. 
\textbf{d}, Empirical MSE obtained with $n=20,000$ noise realisations. $\overline{MSE}$ is its average over all pixels (except the first pixel for one-step multiplexing, see Table. \ref{tab:SumUp_HadaSmatCos})
\textbf{e}, Section of $\mathbf{MSE}(\hat{\mathbf{x}})$ along the anti-diagonal. The dashed-lines represent theoretical MSE values, and the faint strip the error-bar (see Supplementary methods). Inset: ground-truth object of average $\bar{x}$. 
\textbf{f}, Example of one estimate for a sparse object and a 'negative' object (object with structures of interest dimmer than a bright background), for raster-scanning and one-step multiplexing. The associated complete results are shown in Fig S7. 
}
\label{fig:SimuResults}
\end{figure*}

To assess the influence on the object structure, we simulate RS and positive-Hadamard multiplexing on a typical intensity object with beads of different brightness (Fig.~\ref{fig:SimuResults}). 
After a single realisation of the data (Fig.~\ref{fig:SimuResults} b-c), the brightest bead on the top right appears less noisy with multiplexing than with raster-scanning; dual-detection leading to the least noise and two-step multiplexing to the most. 
In opposite, the dimmest bead on the bottom left appears much noisier with multiplexing, to the point that it is nearly buried into the background noise in the two-step scheme. Repeating the simulation $n=20,000$ times leads to an 'empirical' MSE value (Fig.~\ref{fig:SimuResults} d-e). As predicted, the MSE of RS tends towards the ground- truth object $\mathbf{x}$, and the positive-Hadamard multiplexing MSE is constant over all object pixels, approaching the theoretical values of Table.~\ref{tab:SumUp_HadaSmatCos}. 
On the MSE section plot (Fig.~\ref{fig:SimuResults} e) it appears clearly that, as compared to raster-scanning, the three multiplexing schemes degrade the MSE (and SNR) on all pixels along the anti-diagonal, except on the brightest bead. For example, one-step multiplexing, improves the SNR by 1.7 times on the brightest bead, degrades it by the same amount on the dimmest bead and by 4 times on the background. 
This example illustrates a key point to consider when choosing between PHC-multiplexing or raster-scanning: the magnitude of the SNR gain and loss essentially depends on how the object pixels are distributed as compared to the object average signal $\bar{x}$. Two utmost cases are illustrated on Fig.~\ref{fig:SimuResults} f: On a sparse object, the few non-zero pixels of interest are most likely much brighter than $k\bar{x}$: a substantial SNR gain is then expected on those pixels, although it is degraded on null pixels. In opposite, for a 'negative' object (object with structures of interest dimmer than a bright background), the structures of interest are likely to be dimmer than $k\bar{x}$: a SNR loss is then expected on most pixels of interest. The detailed results for these two objects are shown in Fig.~S7 (Supp. Methods)
Note that, since the SNR is degraded on the background, the peak-to-background ratio is systematically worsened as compared to RS, independently of the object structure \cite{ScotteThesis2020}. \\

These results are confirmed experimentally, on an optical system where the noise only arises from the photon-counting process (Fig.~S4 and S5 (Supp. Methods)
Here, to bypass dependence on the experimental ground truth, we do not calculate the MSE but rather the estimation variance $\mathbf{V}$ after $n$ experiments (In the absence of bias, $\mathbf{MSE}=\mathbf{V}$). 
All other detailed experimental methods can be found in the Supplementary Methods. 
First, we consider the case of 2-D fluorescent imaging: we detected the fluorescent signal emitted by fluorescent particles deposited on a glass slide. In essence, the experimental setup 
is similar to the scheme of Fig.~\ref{fig:ConceptFig}a, where the single-pixel detector is a photomultiplier tube operating in photon-counting mode; and the multiplexing matrix is physically implemented onto a digital micromirror device (DMD). This 2-D array of micromirrors - acting as a binary modulator - contains the magnified 2-D fluorescent image (Fig.~\ref{fig:ExperimentalResults} (left inset-c)).
For the comparisons between multiplexing and RS to be reliable, we implement RS directly onto the DMD plane, which is formally equivalent to scanning the sample plane with a point-focus.
The first sample (Fig.~\ref{fig:ExperimentalResults} a-c) is relatively sparse, and the particles are more than 10 times brighter than the sample average signal ($\approx 1.6$ counts). After one experiment, the SNR is visually improved on the particles (Fig.~\ref{fig:ExperimentalResults}a). Repeating the experiment $n=20$ times leads to an approximately constant variance (Fig.~\ref{fig:ExperimentalResults} b-c) that confirms the theoretical values of Table.~\ref{tab:SumUp_HadaSmatCos}. In addition, to study the effect of sparsity, we perform the same experiments on a second sample (same physical object but cropped onto the DMD plane). This sample is not so sparse, and the particles intensity is only about twice higher than the sample average signal ($\approx 9$ counts). This time, dual-detection improves the SNR on the particles and slightly degrades the background, while two-step multiplexing clearly degrades the SNR on all pixels (Fig.~\ref{fig:ExperimentalResults} d-f).
These experimental results confirm the theoretical MSE values for multiplexing, but one may notice that the variance associated with RS is not exactly equal to the object. It comprises an offset due to the imperfect DMD contrast, which impact is negligible for multiplexing, but significant for RS (see Fig.~\ref{fig:Other_BckNoise_Simu}d and Supplementary Methods}). Yet, this artefact only comes from the fact that we mimic RS measurements with the DMD: in practise RS does not involve a multiplexing element but a focussed beam that would not degrade the performance in the same way. \\
Secondly, to highlight that the results of this paper are not restricted to imaging, we also confirm the results on Raman spectroscopy experiments. There, the object $\mathbf{x}$ is a 1-D intensity spectrum (Fig.~\ref{fig:ExperimentalResults}j-right inset), 
and the multiplexed quantities are no longer spatial pixels but wavelength bins of the spectrum (Fig.~\ref{fig:ExperimentalResults}g). 
In the optical setup Fig.~S2 b, 
a sample emits Raman intensity containing several wavelengths, which are dispersed with a diffraction grating. The DMD plane thus contains a 1-D Raman spectrum (Fig.~\ref{fig:ExperimentalResults}j-left inset) which can then be modulated.
The sample is a liquid solvent (Dimethyl Sulfoxide), acquired for two different integration times (5~ms and 2~ms). The variance results obtained with $n=1000$ measurements validate the theoretical values. On this sample, dual-detection is advantageous everywhere but on the background; one-step multiplexing is advantageous everywhere but on the background and dimmest peak; and two-step multiplexing is only advantageous on the brightest peaks. For one random experiment, this may directly result in noisy dim peaks (Fig.~\ref{fig:ExperimentalResults}h) or even in undistinguishable dim peaks (Fig.~\ref{fig:ExperimentalResults}i). 
Note that the two-step multiplexing scheme
is not physically relevant for a 1-D Raman spectrum, but is mimicked with one step multiplexing with an equivalent matrix $\mathbf{M} \otimes \mathbf{M}$ (Fig.~\ref{fig:Modalities_MainTxt}. Also note that here, the impact of the imperfect DMD constrast in negligible (Supplementary Methods).\\

\begin{figure*} [h!]
\centering
\includegraphics[width=\linewidth]{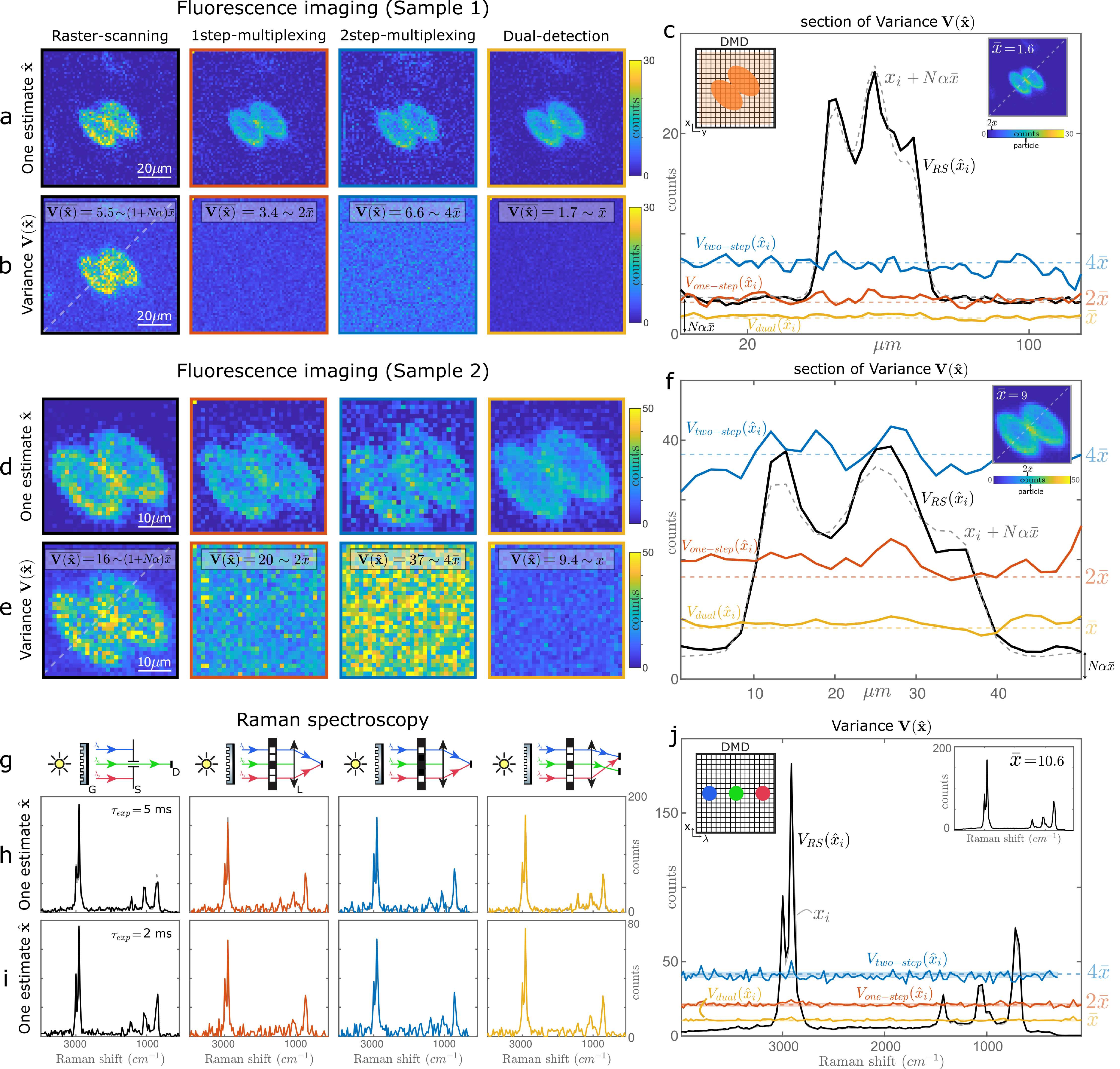} 
\caption{Experimental results of raster-scanning and positive-Hadamard multiplexing in fluorescence imaging (\textbf{a-f}) and in Raman spectroscopy (\textbf{g-j}). \textbf{a}, Example of estimation after one single experiment. \textbf{b}, Estimation variance from $n=20$ experiments, and its average over all pixels $\overline{\mathbf{V}(\hat{\mathbf{x}})}$ (except the 1$^{st}$ pixel for 1step-multiplexing). 
\textbf{c}, Section of the estimation variance along the image anti-diagonal. Dashed-lines: predicted theoretical values of Table \ref{tab:SumUp_HadaSmatCos}. The raster-scanning offset is discussed in the section (Robustess to perturbations). Right inset: object average with mean signal $\bar{x}$. Left inset: Representation of the sample imaged onto the DMD plane. 
$N=4096$: number of object pixels, $\alpha \approx 0.065\%$: accounts for the imperfect DMD contrast, $N\alpha \bar{x} = 4.25$ counts: resulting offset. 
\textbf{d-f}, Same as \textbf{a-c} for Sample 2, with $N=1024$ and $N\alpha \bar{x} = 5.7$ counts.
\textbf{g}, Schematic representation of a dispersive spectrometer in the different modalities. G: diffraction grating, S: slit, D: single-pixel detector, L: converging lens. See Fig.~S2 
b for more details;
\textbf{h-i}, Example of spectrum estimation after one single measurement (5~ms and 2~ms exposure time per spectral bin). 
\textbf{j}, Estimation variance with $\tau_{exp} = 5$ ms from $n = 1000$  experiments with error bars (Supp. Methods). Dashed-lines: predicted theoretical values from Table \ref{tab:SumUp_HadaSmatCos}. 
Right inset: object average with mean signal $\bar{x}$. Left inset: Representation of the the DMD plane containing dispersed wavelengths constituting a spectrum.}
\label{fig:ExperimentalResults}
\end{figure*}

\noindent \textbf{Positive-Cosine multiplexing: }
For positive-cosine multiplexing, the MSE results are analogous to positive-Hadamard multiplexing: the above analyses hold, but the constant $k$ is not the same.
{\renewcommand{\arraystretch}{2}
\begin{table} [H]
\centering
\begin{tabularx}{\linewidth}{|p{0.7cm}|X|X|p{1.1cm}|p{1.1cm}|}
\hline
 &   One-step & Two-step  & \multicolumn{2}{|c|}{Dual-detection}  \\
\hline
\scalebox{0.9}{$\mathbf{A}$} &  \scalebox{0.9}{$\mathbf{M}$} & \scalebox{0.9}{$\mathbf{M}^T \otimes \mathbf{M}$}  & \scalebox{0.8}{$ \left[ \mathbf{M} \hspace{0.1cm} \mathbf{M_2} \right]^T$} & \scalebox{0.8}{$\mathbf{M}-\mathbf{M_2}$} \\ 
\hline \hline
\multicolumn{5}{|c|}{Positive-Cosine multiplexing (\scalebox{0.9}{$\mathbf{M} = \mathbf{C_1}$)}} \\
\hline
\scalebox{0.9}{$MSE$} &  $4 \bar{x}$  $\scriptstyle (\forall i \neq 1 )$  & $16 \bar{x}$  $\scriptstyle (\forall i \neq n_1)$ & \multicolumn{2}{|c|}{$ 2 \bar{x}$   $\scriptstyle ( \forall i )$} \\
\hline
\end{tabularx}
\caption{MSE for positive-cosine multiplexing with the matrix $\mathbf{C_1}$ defined in equation \eqref{eqn:C1DCT_def} - for the three multiplexing schemes of Fig.~\ref{fig:Modalities_MainTxt}. Results hold for LS-estimation and large number of pixels $N \gg 1$. Results for other forms of positive-cosine multiplexing are detailed in \cite{Scotte2022}, section 4.3 
}
\label{tab:SumUp_Cos}
\end{table}
Table.~\ref{tab:SumUp_Cos} and Fig.~S8 (Supp. Methods) 
give the values of $k$ for positive-cosine multiplexing with the specific positive DCT matrix defined in 
equation \eqref{eqn:C1DCT_def}. 
They show that, one-step multiplexing leads to a MSE equals to four times the object average ($k=4$). Comparatively, two-step multiplexing squares the MSE ($k=16$); and dual-detection improves it by a factor two ($k=1$). Here again, one-step multiplexing leads to a better MSE than two-step multiplexing, and dual-detection improves the MSE. 
These result also show that positive-cosine multiplexing with the matrix $\mathbf{C_1}$ consequently degrades the MSE as compared to positive-Hadamard multiplexing. This SNR loss is particularly visible in the two-step scheme (Fig.~S8 (Supp. Methods))
, where positive-cosine multiplexing is worse than raster-scanning on all beads - including the brightest one - and where the dimmest bead (bottom left) is completely buried into the background noise. Overall, a similar analysis as for Fig.~\ref{fig:SimuResults} can be drawn, but positive-cosine multiplexing does not improve the SNR of RS on as many pixels as with positive-Hadamard multiplexing. 
Note that the results of Table.~\ref{tab:SumUp_Cos} results are partially empirical: we prove theoretically (\cite{Scotte2022}, section 4.3) that the MSE of a general positive-cosine multiplexing scheme is constant for $N \gg 1$, but the values of $k$ for the specific matrix $\mathbf{C_1}$ matrix are deduced from simulations 
(Fig.~S8 Supp.Methods). \\

\noindent We emphasize that these values of $k$ are not general for all forms of positive-cosine multiplexing, but are only valid for the matrix $\mathbf{C_1}$ defined in equation \eqref{eqn:C1DCT_def}. Here, as for the matrices $\mathbf{H_1}$ and $\mathbf{S}$, the matrix coefficients are comprised between 0 and 1. By setting this constraint, we chose the point-of-view of a user of a typical incoherent optical system, where the modulation possibilities are often comprised in this range. 
Positive-cosine modulation can be performed in other manners, potentially leading to different values of $k$. Diverse positive-matrices built upon the discrete cosine transform can be used (e.g. with other normalisation factor); and the system architecture by itself can define a different multiplexing matrix \cite{Futia2011, Scotte2019}. If the multiplexing matrix is simply proportional to $\mathbf{C_1}$, the MSE would be modified according to Table.~\ref{tab:PertubEpsilon}. 
Note that positive-cosine modulation also applies to cases where multiplexing is achieved via interferometric measurements (e.g. Fourier-transform infrared spectroscopy) \cite{Fellgett1967, Harwit1979, Bialkowski1998, Fuhrmann2004}. Such systems do not fall into the scope of this text because the modulation does not happen in intensity. Yet, they comply with the model of equation \eqref{eqn:completemodel}, where the field power spectrum (object $\mathbf{x}$) is linearly related via some positive cosine transform to the measurements \cite{1385, Fuhrmann2004}. We show in \cite{Scotte2022} that the results of the present text also apply to such interferometric systems, to a constant. They also lead to a constant MSE and comply with equation \eqref{eqn:MSE_kx}).  Last, note that for positive-cosine multiplexing, there exist alternative solutions to the dual-detection scheme that are likely to further improve the MSE - such as the common four-step phase-shifting method \cite{Zhang2015, Meng2020}. Such strategies are not considered in this work. \\

\noindent \textbf{Implications: }
First, our results highlight that, for PHC-multiplexing, the SNR is substantially affected by the system design: when possible, one-step multiplexing should be preferred over the two-step scheme, and implemented in a dual-detection mode. Second, they show that positive-Hadamard multiplexing should be preferred over positive-cosine multiplexing with the matrix $\mathbf{C_1}$. Last, they feature that the benefit of PHC-multiplexing over raster-scanning depends on how the intensity is distributed over the object pixels: PHC-multiplexing is mostly beneficial for samples which features of interest are brighter than $k$ times its average value $\bar{x}$: in particular, it can be present a great advantage on sparse objects, but should be avoided on "negative" objects. These results are not particularly intuitive and contrast with the multiplexing advantage that holds under additive white Gaussian noise (Fig.~\ref{fig:ConceptFig} b). There, the noise is independent of the signal, thus more signal comparatively means less noise. With photon-noise, the key point is that the noise is \textit{depends} on the signal: the noise variance scales with the detected signal (replacing the Poisson noise with Gaussian noise of variance equal to the signal lead same results). Then, in raster-scanning, the photon noise on each pixel is associated with its own brightness, and a null pixel does not induce photon noise. 
In opposite, PHC-multiplexing combines photons from object parts of different brightness, collecting a large signal varying about a high positive DC value (Fig.~\ref{fig:ConceptFig} a). When demultiplexing, the large noise associated with this DC value seems to spread over the whole object, thereby risking to bury the signal of a faint pixel into the photon-noise of bright ones \cite{Studer2012, Garbacik2018}.

\subsection*{Impact of other estimators}
\label{section:Algos}

\begin{figure*} [hbt!]
\centering
\includegraphics[width=\linewidth]{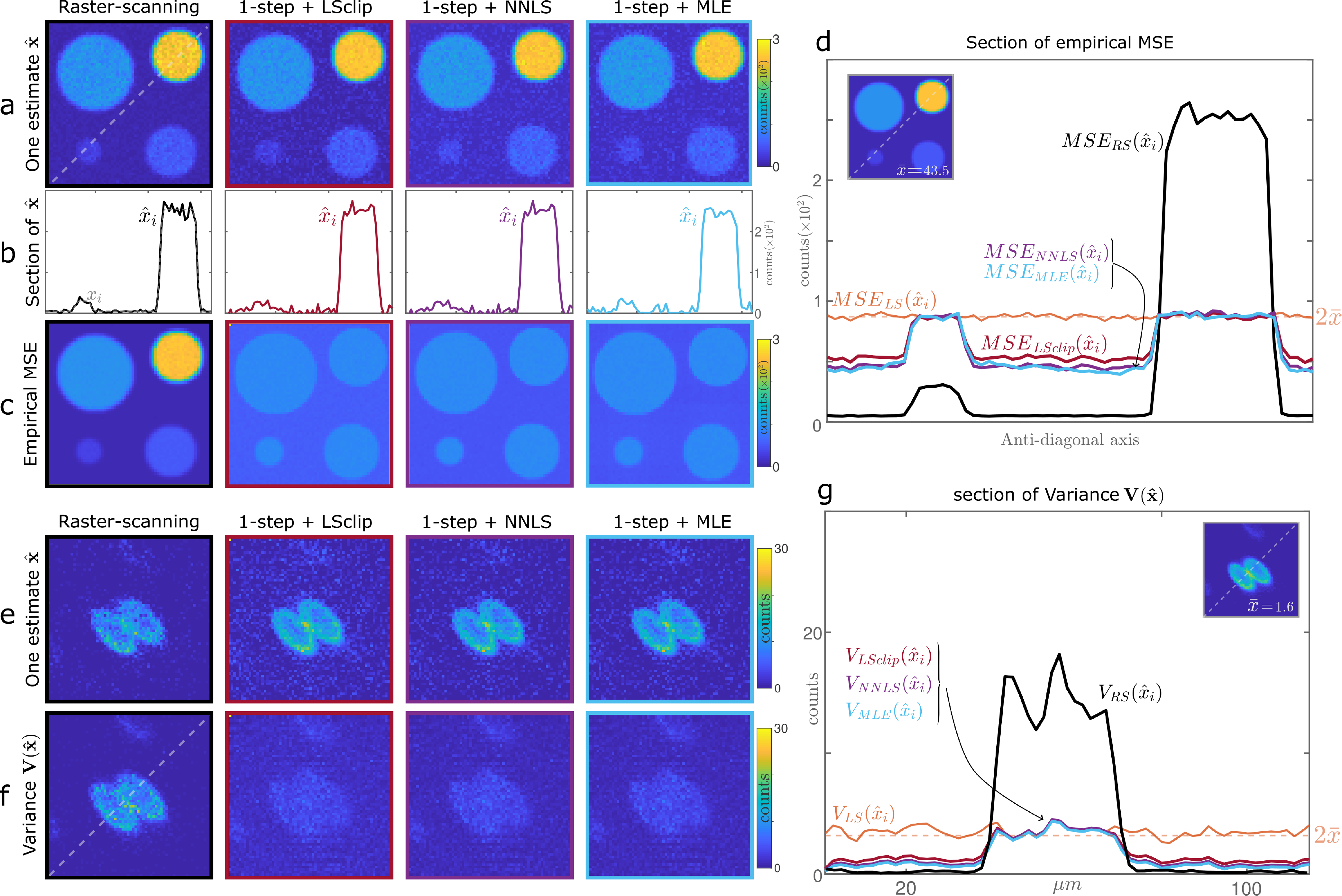} 
\caption{Effect of the proposed estimators, for one-step positive-Hadamard multiplexing, as compared to LS-estimation (Fig.~\ref{fig:SimuResults}-\ref{fig:ExperimentalResults}). \textbf{a-d,} Results for the simulated object of Fig.\ref{fig:SimuResults}, with $n= 5000$ realisations. The estimators reduce the MSE on the background. \textbf{e-g,}
Experimental results for fluorescence imaging, as described in Fig.~\ref{fig:ExperimentalResults}, with $n= 20$ realisations. The estimators reduce the MSE on the background. Here the MLE is applied to raster-scanning and improves the original offset (Supplementary Methods).}
\label{fig:Effect_estimators_Allobjects}
\end{figure*}

{\renewcommand{\arraystretch}{2}
\begin{table*} [htb!]
\centering
\begin{tabularx}{\linewidth}{|p{2.7cm}|p{3.5cm}|X|p{2.7cm}|}
\hline
Pixel intensity & Estimator & SNR of PHC-multiplexing as compared to raster-scanning ? & SNR gain / loss \\
\hline
$x_i \geq k \bar{x}$ & All & SNR improvement  & gain $= \sqrt{ \frac{x_i}{k\bar{x}} }$ \\
\hline
$x_i \leq k \bar{x}, x_i \neq 0$ & LS & SNR degradation & loss $=\sqrt{ \frac{x_i}{k\bar{x}} }$\\
 & LSclip, NNLS, MLE & SNR degradation but the degradation is mitigated on dim parts &  loss $\leq \sqrt{ \frac{x_i}{k\bar{x}} }$ \\
\hline
$x_i \leq k \bar{x}, x_i = 0$ & LS & SNR degradation & loss $=\sqrt{ \frac{x_i}{k\bar{x}} }$  \\
& LSclip & SNR degradation but mitigated  & loss $ \leq  \sqrt{ \frac{x_i}{k\bar{x}} }$ \\
& NNLS, MLE  & $\approx$ no SNR modification & $\approx $ no change  \\
\hline 
\end{tabularx}
\caption{Indicative effect of the estimators on the threshold value of equation \eqref{eqn:ruleofthumb_kx}. LS: least-square, LS-clip: least-square with positive threshold, NNLS: non-negative least-square, MLE: maximum-likelihood estimator (definitions in Supplementary Methods). \textit{This indicative table is valid except for extremely dim samples where the positively constrained estimators may impact even pixels higher than $k \bar{x}$.}}
\label{tab:SumUp_Ruleofthumb_Algos}
\end{table*}

The above results are valid when the object is retrieved via least-square estimation. Yet, this estimator does not take into account some \textit{a priori} knowledge such that (i) the object is a positive quantity (ii) nor the nature of the Poisson noise. Therefore, we consider three simple alternative estimators: a LS-estimator with positive threshold (\textit{LS-clip}); an estimator that takes into account the positivity constraint (\textit{NNLS}: non-negative least-square); and an estimator that take into account both the positivity constraint and the nature of the noise (\textit{MLE}: Maximum Likelihood Estimator). Details on these estimators are provided in Supplementary Methods. Here, the aim is not to provide a complete study, but rather to identify in which cases they may be useful to improve the MSE. 
The simulations of Fig.~\ref{fig:Effect_estimators_Allobjects} and Fig.~S9 
assess the performance of these estimators on different types of samples. 
Essentially, they show that these estimators mostly reduce the MSE on object pixels where the positivity constraint can be enforced (i.e. on lowest-intensity or zero-valued object pixels), and that the dimmer the pixel, the more NNLS and MLE are beneficial over LS-clip. 
This is clear on the MSE of Fig.~\ref{fig:Effect_estimators_Allobjects}d,g: as compared to LS, the MSE is reduced on the background but not on the particles. On the estimate, this translates to a reduction of the background noise, and the dimmer the pixels, the stronger is the noise reduction. On Fig.~\ref{fig:Effect_estimators_Allobjects}a,e, the effect is quite visible on the background, but is most pronounced when the object is sparse (Fig.~S9 a) 
In Supplementary Methods, we show that the MLE seems to perform better than NNLS at reducing the MSE on the background without introducing a consequent bias in the estimation (Fig.~S1) 
, but this is at the expense of computational complexity. Note that these algorithms do not include a sparsity-prior, the error reduction is simply due to the positivity constraint.\\

\noindent Overall, the considered estimators do not necessarily bring an improvement over LS (eg. in
Fig.~S9 b). 
They are mostly beneficial for sparse objects (Fig.~S9 a)
or objects with dim parts (Fig.~\ref{fig:Effect_estimators_Allobjects}). LS-clip improves the MSE by discarding potential negative estimated values, and NNLS and MLE bring an additional improvement if the object is sparse or comprises null pixels.
In any case, as summarized in Table.~\ref{tab:SumUp_Ruleofthumb_Algos}, equation \eqref{eqn:ruleofthumb_kx} remains globally valid: PHC-multiplexing brings a SNR improvement over raster-scanning for pixels brighter than $k\bar{x}$. For dim pixels under this threshold value, these estimators can partially mitigate the SNR degradation. Yet, it is in the presence of null pixels that estimators such as MLE or NNLS are most useful: on these pixels, they can completely counterbalance the SNR degradation induced by the use of PHC-multiplexing with LS-estimation, which is particularly useful for sparse objects.



\subsection*{Robustness to perturbations} 
\label{sec:otherModelsBck}

Experimentally, several noise sources - such as the ones depicted on Fig.~\ref{fig:block_diagram} - may sometimes perturb the initial photon-noise limited system of equation~\eqref{eqn:completemodel}. Therefore, it is important to assess which of RS or PHC-multiplexing is most robust to system perturbations. Here, we study their robustness to: (i) additional electronic noise $\mathbf{e}$ arising from the detector, (ii) additional signal $\bm{\eta}$ entering the system after the multiplexing step, (iii) additional signal $\bm{\beta}$ entering the system before the multiplexing step, (iv) a constant offset $\alpha$ in the multiplexing matrix itself. 
\begin{figure} [H]
\centering
\includegraphics[scale=0.6]{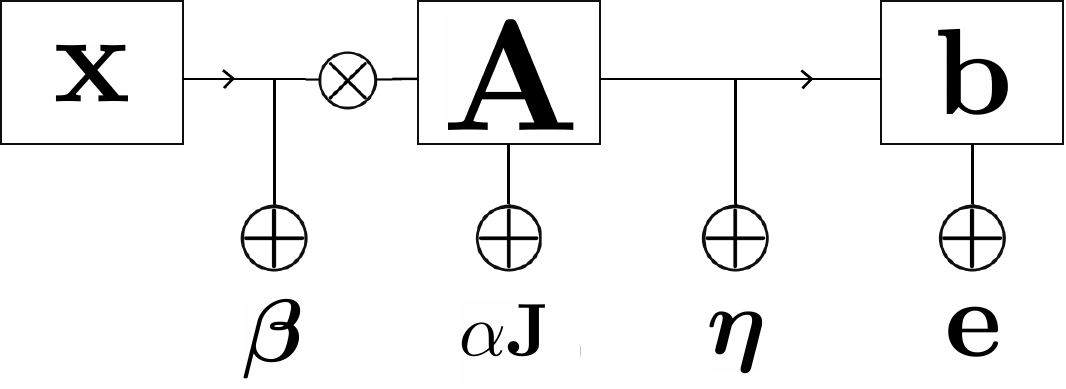} 
\caption {Additional nuisance sources. $\mathbf{e}$: additive white Gaussian noise of variance $\sigma^2$. $\bm{\eta}$, $\bm{\beta}$: unwanted signal adding to the object signal. $\alpha$: constant offset added to the multiplexing matrix, $\mathbf{J}$: the matrix of ones. $\bm{\eta}$, $\bm{\beta}$ and $\alpha$ are real positive quantities and assumed to be known from a calibration step.}
\label{fig:block_diagram}
\end{figure}
The theoretical results \cite{Scotte2022}, section 5.2) and simulations (Fig.~\ref{fig:Other_BckNoise_Simu}) show that, when the number of pixels is sufficiently large, PHC-multiplexing is robust to these additional perturbations, except when the unwanted signal $\bm{\beta}$ undergoes multiplexing (Fig.~\ref{fig:Other_BckNoise_Simu}c). Conversely, RS is not robust to these perturbations, since the noise variance or magnitude adds as an offset to the MSE (Fig.~\ref{fig:Other_BckNoise_Simu}a-d). 
Hence, PHC-multiplexing is more robust than RS to additional signal independent noise $\mathbf{e}$, to unwanted non-multiplexed signal $\bm{\eta}$, and to a multiplexing offset as $\bm{\alpha}$. In these cases, the initial equation \eqref{eqn:ruleofthumb_kx} is lowered by an amount proportional to the strength of the nuisance: the larger the nuisance signal, the more pixels benefit from PHC-multiplexing. 
However, in the presence of an unwanted multiplexed signal $\bm{\beta}$, PHC-multiplexing is less robust than RS, since its MSE is on average $k$ times more impacted. There, the more nuisance, the more pixels benefit from raster-scanning. 

\begin{figure} [H]
\centering
\includegraphics[width=\linewidth]{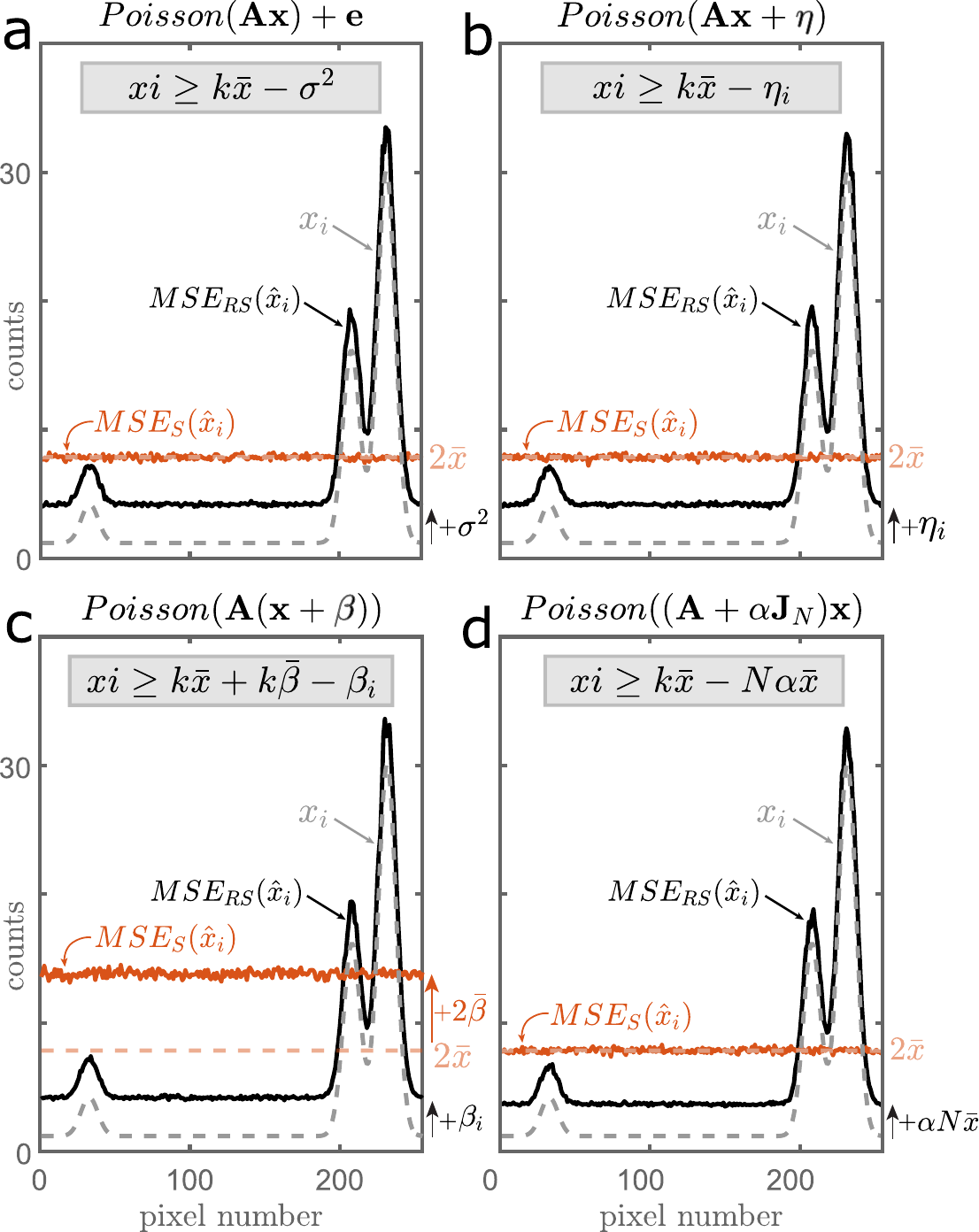} 
\caption {Impact of the nuisance sources on the MSE. Top: noise model. Grey box: PHC-multiplexing improves the SNR on pixels $i$ brighter than the indicated value. black: MSE associated with raster-scanning; red: MSE associated with positive-Hadamard multiplexing ($\mathbf{S}$-matrix, LS-estimation); dashed-lines: MSE for the initial photon-noise model of equation \eqref{eqn:completemodel}. $\forall i$, $\sigma^2 = \eta_i = \beta_i \approx N \alpha \bar{x} \approx 3$ counts.}
\label{fig:Other_BckNoise_Simu}
\end{figure}
Note that the last scenario (constant offset $\alpha$ on the multiplexing matrix) explains the impact of the imperfect DMD contrast on the experimental results of (Fig.~\ref{fig:ExperimentalResults}). Indeed, we measured that DMD pixels in the 'OFF' order contribute to an amount $\alpha \approx 0.065\%$ to the detected signal. This contribution seems insignificant, but substantially degrades the MSE of raster-scanning with an offset of $\alpha N \bar{x} \approx 5$ counts for fluorescence imaging (Fig.~\ref{fig:ExperimentalResults}a-b) and $\alpha N \bar{x} \leq 1$ count for Raman spectroscopy (Fig.~\ref{fig:ExperimentalResults}c), see details in Supplementary Methods. As for PHC-multiplexing, we notice no effect. Yet, we emphasize that this SNR degradation only comes from the fact that we mimic RS measurements with a DMD: it would not happen in practise since RS does not involve a multiplexing device. It is nevertheless interesting that this minor contrast imperfection most likely explains why it is probably rare to observe a clear advantage for raster-scanning when performing imaging on a DMD, especially if the contrast imperfection is not removed from the raw data (Fig.~S10)
. \\

Another sort of modification of the initial model could be that the initial multiplexing matrix is multiplied by a constant. Then the resulting MSE of PHC-multiplexing is modified according to Table.~\ref{tab:PertubEpsilon} (see details in \cite{Scotte2022}, section 5.1).

{\renewcommand{\arraystretch}{2}
\begin{table} [H]
\centering
\begin{tabularx}{\linewidth}{|p{1.2cm}|p{1.7cm}|p{2.2cm}|X|}
\hline
 &  One-step & Two-step  & Dual-detection\\
\hline
$\widetilde{\mathbf{A}}$ &  $\frac{1}{\epsilon}\mathbf{M}$ & $\frac{1}{\epsilon_1}\mathbf{M} \otimes \frac{1}{\epsilon_2} \mathbf{M}$  & $\frac{1}{\epsilon} \left[
\mathbf{M} \hspace{0.1cm} \mathbf{M_2}
\right]^T$ \\ 
\hline
$\widetilde{MSE}$ &  $\epsilon$ $k\bar{x}$ & $\epsilon_1 \epsilon_2$ $k\bar{x}$   & $\epsilon$ $k\bar{x}$   \\
\hline
\end{tabularx}
\caption{$\widetilde{\mathbf{A}}$: modified multiplexing matrix; $\epsilon$: positive constant;  $\widetilde{MSE}$: resulting MSE for PHC-multiplexing}.
\label{tab:PertubEpsilon}
\end{table}

\subsection*{Impact of a constant number of photons} 
\label{sec:NbPhotonsCst}

Last, we emphasize that we have compared positive-multiplexing and raster-scanning at fixed irradiance and integration time, i.e. when 
the number of photons was \textit{not} constant. In this case, we have shown that even though PHC-multiplexing detects about $N/2$ times more photons than RS, it does not necessarily improves the final SNR. It then seems trivial that, if the number of photons collected by PHC-multiplexing is lowered to be equal to the number of photons detected with RS (e.g. by lowering the laser power), its SNR will be further degraded. In \cite{Scotte2022} section 5.1, we show the MSE is worsened by a factor $N/2$:
\begin{equation}
MSE_{PHC}(\hat{x}_i) \approx \frac{N}{2}k\bar{x} 
\label{eqn:CtePhotons}
\end{equation}
Then, if $N \gg 1$, $G_i \approx 0$, meaning that PHC-multiplexing degrades the SNR on virtually all object pixels. On average, the SNR loss is proportional proportional $\sqrt{N}$.
Therefore, when the measurements are only limited by photon-noise, the common argument (that holds for additive-white gaussian noise) stating that \textit{since PHC-multiplexing allows to detect more photons than raster-scanning, the integration time or laser power can be lowered to obtain the same SNR} is not valid. An illustration of this effect is provided in Fig.~S10. 






\subsection*{Conclusion} 
\label{sec:ConclusionDiscussion}

\begin{figure*} [b!] 
\centering
\includegraphics[scale=0.6]{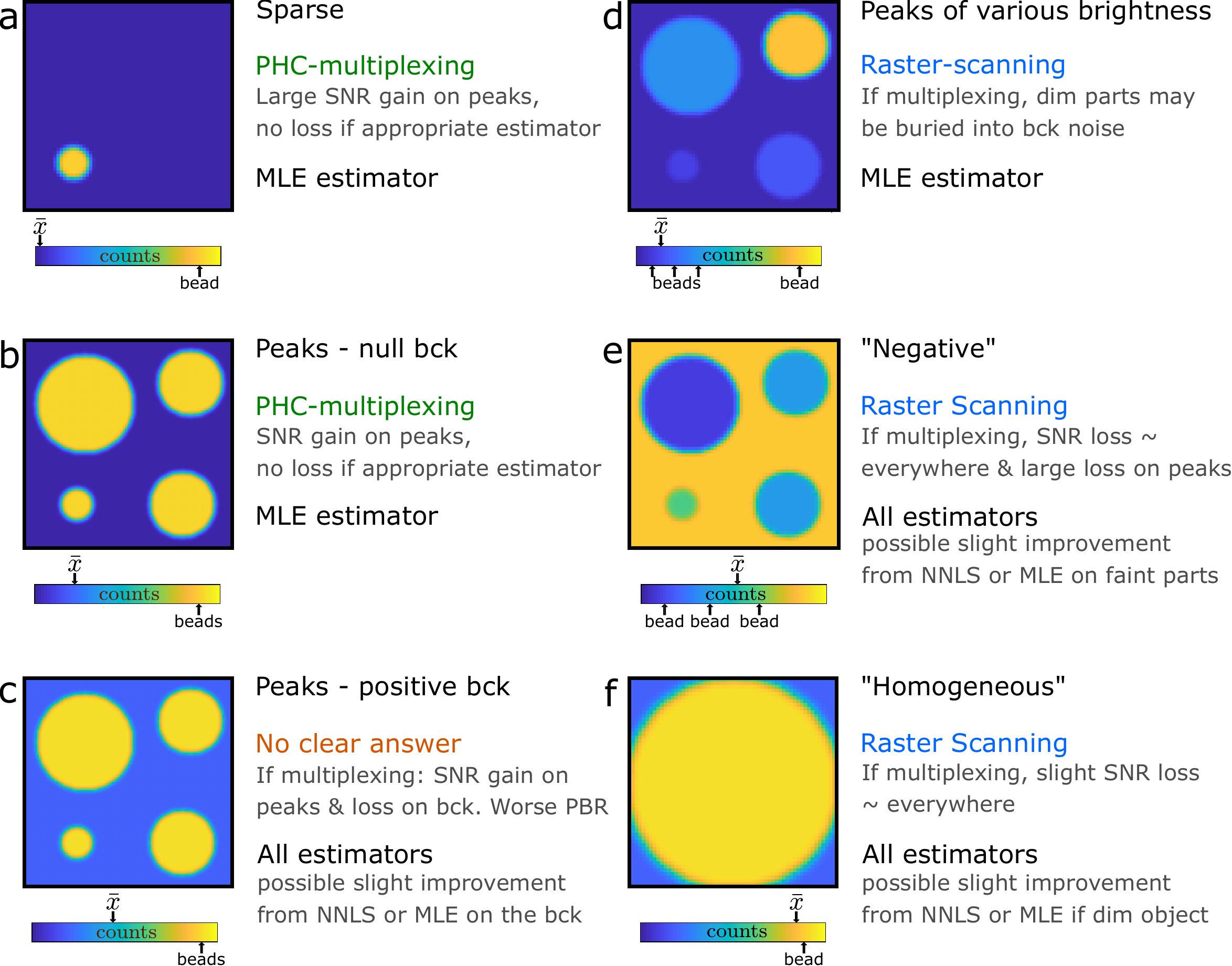} 
\caption{Indicative guideline for the preferred choices between PHC-multiplexing and raster-scanning, for several classes of objects, provided the SNR is the figure-of-merit to optimise.
The preferred estimator choice amongst LS, LS-clip, NNLS and MLE is also indicated. The scale bar indicates the position of the object average signal $\bar{x}$ as compared to the structures of the object. bck: background, PBR: peak-to-background ratio.}
\label{fig:FinalGuideline}
\end{figure*}

This paper compared the SNR of raster-scanning and positive-multiplexing based on Hadamard and Cosine modulation, at fixed integration time and irradiance, when the noise only arises from the photon-counting process. In this context, although PHC-multiplexing detects about $N/2$ times more photons than RS, it does not necessarily improve the SNR of the estimated object. Instead, we showed that the MSE is approximately equal to a constant $k\bar{x}$, meaning that PHC-multiplexing improves the SNR only on pixels at least $k$ times brighter than the object mean signal $\bar{x}$. On pixels lower than $k\bar{x}$, PHC-multiplexing degrades the SNR, except on zero-valued pixels, where the degradation can be mitigated with appropriate estimators. The constant $k$ is at the core of this work: it depends on the multiplexing matrix and on the specific multiplexing configuration. \\

\noindent These results highlight that, for PHC-multiplexing, the SNR is substantially affected by the system design: when possible, one-step multiplexing should be preferred over the two-step scheme, and implemented in a dual-detection mode. Indeed, as compared to one-step multiplexing, the two-step scheme squares $k$ and dual-detection divides it by two. For dual-detection, we also showed that the balanced-detection strategy can be used for the matrices $\mathbf{H_1}$ and $\mathbf{C_1}$, but should not be used for the $\mathbf{S}$-matrix. 
They also show that positive-Hadamard multiplexing leads to a better SNR than positive-Cosine multiplexing with $\mathbf{C_1}$, although this may differ for other types of positive-cosine multiplexing modulations. \\
Most importantly, these results highlight that the benefit of PHC-multiplexing over raster-scanning depends on how the intensity is distributed over the object pixels, i.e. on the object structure.  Therefore, the question: \textit{Does PHC-multiplexing leads to a better SNR than raster-scanning for photon-noise limited data?} has no straightforward universal answer. The results depends on the type of sample and on the user's interest. Yet, we provide an indicative guideline table with typical sample types and the preferred estimator to use (Fig.~\ref{fig:FinalGuideline}), when the SNR is the figure-of-merit to optimize. Altogether, raster-scanning should be preferred when pixels of interest lie under $k\bar{x}$, such as in homogeneous objects, "negative" objects, or objects with structures of very different brightness. Conversely, PHC-multiplexing should be preferred for objects with some large intensity parts on a faint or null background; and finds its greatest advantage for sparse objects. \\

\noindent With this study, we hope to have clarified a few crucial points concerning the choice of some acquisition strategies and their signal-to-noise ratio. Yet, it also leaves many open questions. First, concerning the validity framework of the results. In this text, we focused on intensity modulation multiplexing, for positive-Hadamard and Cosine modulation. We showed \cite{Scotte2022} that the results also hold for some systems where the modulation does not happen in intensity (positive-cosine-multiplexing via interferometric measurements such as in Fourier-transform infrared spectroscopy), consistently with \cite{Fellgett1967, Harwit1979, Bialkowski1998, Fuhrmann2004}.
It is also possible that our results hold for intensity modulation multiplexing with other deterministic real positive matrices: in \cite{Scotte2022}, we give some theoretical conditions on such matrices that may help to answer this question. \cite{Scotte2022} also provides a detailed methodology with general results to facilitate the adaptation to other multiplexing matrices.
In addition, it would also be of great interest to conduct a similar SNR analysis for non-deterministic modulations, for example with speckle intensities or positive random matrices \cite{Raginsky2010, Marcia2011, Liutkus2014, Guerit2021}. 
Another important aspect to consider is the impact of the number of measurements. Indeed, one advantage of positive-multiplexing is that it can be applied to undetermined systems with techniques such as compressive sensing \cite{Candes2006a, Davenport2012}. But there also, it is crucial to identify the correct noise hypothesis that may impact the performances of some widely used computational methods \cite{Raginsky2009, Willett2021}.
Otherwise, many other parameters could be investigated to complete our SNR study: One could for instance apply the same study to non-linear systems \cite{Audier2020, Heuke2020}, or consider the impact of the resolution and sampling \cite{Sun2016, Sha2020}; of other estimation methods with sparsity priors \cite{Ratner2007a}; or of more complex sources of noise \cite{Nitzsche2003}.


\cleardoublepage

\noindent \textbf{Funding}\\
C. S. has received funding from the H2020 Marie Skłodowska-Curie Actions (713750). This research has received funding from EU ICT-36-2020RIA CRIMSON, Agence Nationale de la Recherche  (ANR-21-ESRS-0002 IDEC), Centre National de la Recherche Scientifique, Aix-Marseille University. \\

\noindent \textbf{Acknowledgements}\\
The authors thank Simon Labouesse, Siddharth Sivankutty, Philippe Réfrégier, Laurent Jacques, Randy A. Bartels, Marc Allain, Anne Sentenac, Sandro Heuke and Luis Arturo Aleman Castaneda for fruitful scientific discussions. \\ 

\noindent \textbf{Authors Contributions}\\
C.S. performed the calculations, simulations and experiments, and wrote the paper. All authors contributed to the scientific discussion and revision of the paper.\\

\noindent \textbf{Competing interests}\\
The authors declare no conflict of interest.\\

\noindent \textbf{Additional information} \\
This paper is supported  by a Supplementary Information that provides the detailed theoretical proofs and derivation methodology, available at : \url{https://arxiv.org/abs/2204.06308}.

\cleardoublepage


\renewcommand{\thefigure}{S\arabic{figure}}
\setcounter{figure}{0}

\section*{Supplementary Methods}

\subsection*{Estimators}
\label{sec: Description of the estimators}

\subsubsection*{Least-square estimation (LS)}
\label{sec: LS estimator}
The LS estimator minimizes the squared $l_2$ norm between the noisy and noiseless measurements. The LS solution reads:
\begin{equation} 
\mathbf{\hat{x}}_{LS} = argmin ||\mathbf{b}-\mathbf{b_0}||^2 = \mathbf{A}^{-1} \mathbf{b}
\label{eqn:LSsol}
\end{equation}
if $\mathbf{A}$ is invertible.
The LS estimator is optimal in the sense of the Maximum-Likelihood for AWGN. Under Poisson noise with no constraint on the estimate $\mathbf{\hat{x}}$, and if $\mathbf{A}$ is invertible, the LS estimate is efficient, meaning is unbiaised with variance equal to the Cramer-Rao lower bound \cite{Fuhrmann2004, Palkki2009, Refregier2018}. Yet, here, the objects of interest are positive intensities and the measurements number of photons counts. We thus consider in the following estimators with positivity constraints. 

\subsubsection*{Least-square estimation with negative values removal (LS-clip)}
\label{sec: LSclip estimator}
The simplest method to take into account the positivity of the object is to find the LS estimate \eqref{eqn:LSsol} and set the negative values of $\mathbf{\hat{x}}$ to zero. We call this \textit{ad hoc} method LS-clip. We choose to include this method because it reflects the commonly applied positive threshold on experimental results. 

\subsubsection*{Non-negative Least-square estimation (NNLS)}
\label{sec: NNLS estimator}
The NNLS estimator takes into account the positivity of the object by solving the LS problem with positivity constraints: 
\begin{equation} 
\label{eqn:NNLSsol}
\mathbf{\hat{x}}_{NNLS} = arg min \ ||\mathbf{b}-\mathbf{b_0}||^2 \ \text{subject to} \ x_i	\geq 0
\end{equation}
For simplicity, we use the in-built Matlab function \textit{lsqnonneg} based on \cite{Lawson1995}. On the studied objects, we verified that it approximately behaves as FISTA with positivity constraints.

\subsubsection*{Poisson Maximum-likelihood estimate with positivity constraints (MLE)}
\label{sec: MLE estimator}
To better take into account the photon noise model, we use an estimator derived from the Poisson distribution. 
For statistically independent measurements, the probability of observing a particular vector of photons counts $\mathbf{b}$ for a given $\mathbf{x}$  - is given by \cite{Refregier2004, Palkki2009}: 
\begin{equation} 
P(\mathbf{b} ; \mathbf{x}) = \prod_{i=1}^{M} e^{-[\mathbf{Ax} + \mathbf{g}]_i} \frac{([\mathbf{Ax} + \mathbf{g}]_i)^{b_{i}}}{b_{i}!} 
\label{PoissonLikelihood}
\end{equation}
where $\mathbf{Ax}+\mathbf{g} = b_{0i} = \langle b_i \rangle$. Here we add a small constant vector $\mathbf{g} \approx 10^{-3}\mathbf{1}_N$ counts to the initial model 
in order to avoid singularities in the following algorithms. 
$P(\mathbf{b} ; \mathbf{x})$ is called the likelihood for a Poisson distribution. We seek the values of $x_n$ than maximize the likelihood to obtain $b_i$ photon counts, under the positivity constraint  $x_n \geq 0$ ($n = 1..N$). In other words, given $\mathbf{b}$, we seek the maximum-likelihood estimate (MLE)
\begin{equation}
\mathbf{\hat{x}}_{MLE} = arg max \ P(\mathbf{b} ; \mathbf{x}) \ \text{subject to} \ x_i \geq 0
\label{eqn:MLEPoiss}
\end{equation}
To solve the above equation, we use two different algorithms. 
First, we use the expectation–maximization (EM) algorithm (known as Richardson-Lucy algorithm) \cite{Gurioli2014, Lucy1974, Shepp1982}, that searches for the solution of \eqref{eqn:MLEPoiss} by solving
\begin{equation}
\mathbf{A}^T diag(\mathbf{Ax} + \mathbf{g})^{-1}\mathbf{b} - \mathbf{A}^T\mathbf{1} = \mathbf{0}
\label{eqn:PoissonloglikGradient}
\end{equation}
iteratively through: 
\begin{equation}
\mathbf{\hat{x}}^{q+1} = \frac{\mathbf{A}^T diag(\mathbf{A}\mathbf{x}^q + \mathbf{g})^{-1}\mathbf{b}}{\mathbf{A}^T\mathbf{1}}  \odot \mathbf{\hat{x}}^q
\end{equation}
The algorithm is well-established, widely used and simple to implement. It was shown to converge towards a MLE estimation, but there is no guarantee that the maximum is a global maximum \cite{McLachlan2007}.
We initialize the algorithm with the NNLS estimate with an offset given by $\mathbf{g}$. \\
To double-check our implementation of the EM algorithm and its behaviour, we also solve \eqref{eqn:PoissonloglikGradient},with a second algorithm called 'SPIRAL-TAP' (Sparse Poisson Intensity Reconstruction ALgorithms)\cite{Harmany2012b}. 
This algorithm was shown to be stable and converge \cite{Harmany2012b}. In this work we do not include the sparsity constraints that can be taken into account in this algorithm. In all the results of this paper, the two algorithms converge to the same solution, therefore we only show the results for SPIRAL-TAP. 


\subsubsection*{Effect of estimators on the variance and bias}
\label{sec: Effect of the estimators - Variance Bias}
The MSE combines the variance and bias through $MSE = Var + \langle \delta \hat{\mathbf{x}} \rangle ^2$, where the bias is the expected value of the estimation error $\delta \hat{\mathbf{x}}$.
In this section, we empirically study the effect of the different estimators on the estimation variance and bias for three simulated objects. 
Fig.~\ref{fig:dessin_VarBiasAlgos} confirms that the LS-estmator is unbiased, and shows that the MSE is mostly dominated by the effect of the variance. The constrained estimators only reduce the variance where the positivity constraints apply, but this variance reduction can be at the expense of a slight bias (Fig.~\ref{fig:dessin_VarBiasAlgos} b,d).
For the object of Fig.~\ref{fig:dessin_VarBiasAlgos}b, LS-clip trivially adds a significant positive bias on pixels with low or zero value (e.g. 34$\%$ relative error on the background). NNLS and MLE also overestimate the background and both slightly underestimate brighter pixels (few $\%$ relative error). For the sparse object (Fig.~\ref{fig:dessin_VarBiasAlgos} d), the MLE estimator introduces significantly lower bias than NNLS. However, it is well known that MLE introduces some artefacts on edges \cite{Snyder1987}, see the marked pixels (*) . 
These results are consistent with \cite{Vio2004, Fuhrmann2004}. 

\begin{figure*} [hbt!]
\centering
\includegraphics[width=\linewidth]{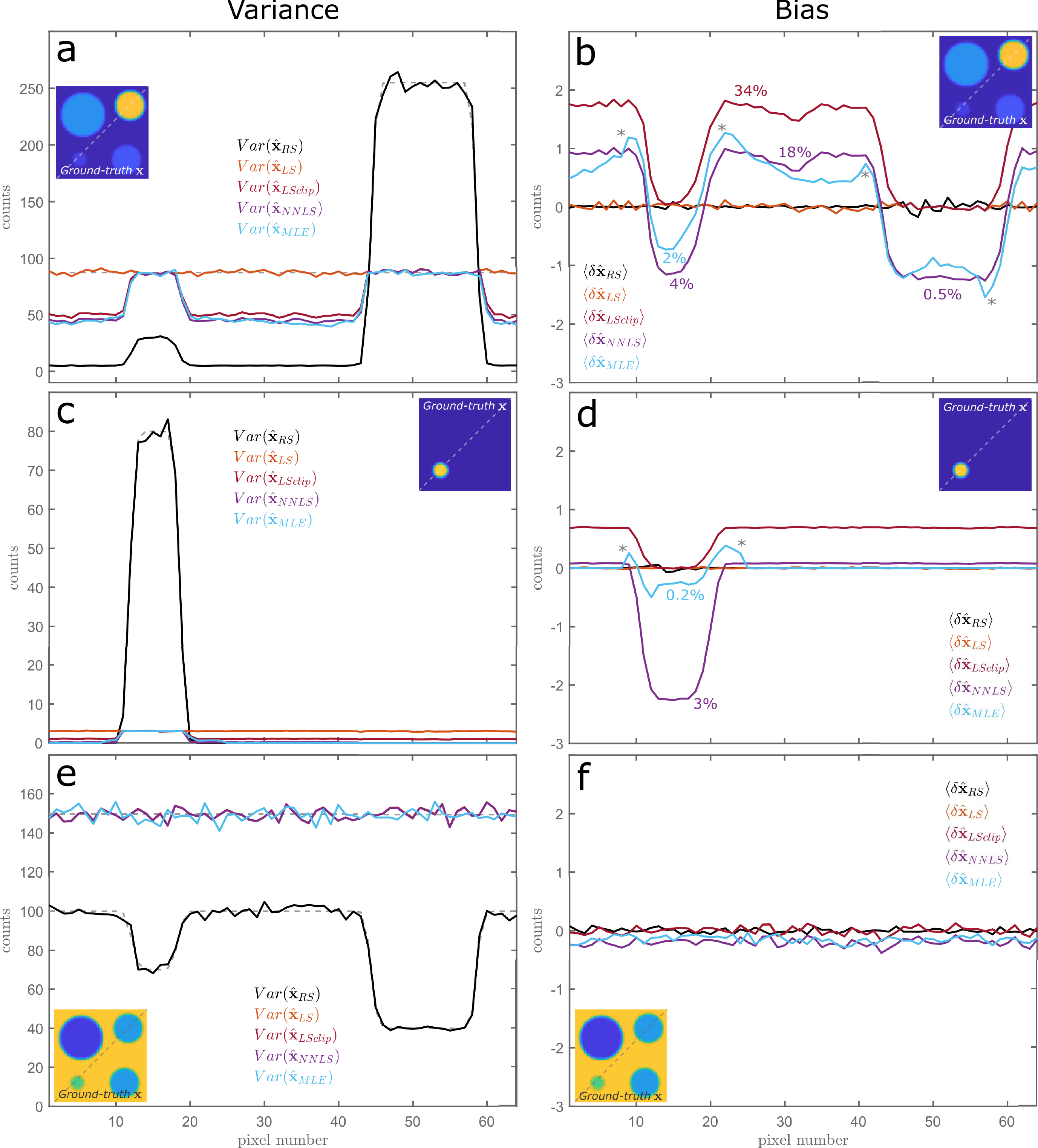} 
\caption{Variance and bias sections along the anti-diagonal associated with the MSE of Fig.~ \ref{fig:SI_Effect_estimators_Allobjects} and Fig.~6 
$n=$5,000 realisations.}
\label{fig:dessin_VarBiasAlgos}
\end{figure*}

\subsection*{Experimental methods}
\label{sec: SI_Experiments}
\begin{figure*} [ht]
\centering
\includegraphics[width=\linewidth]{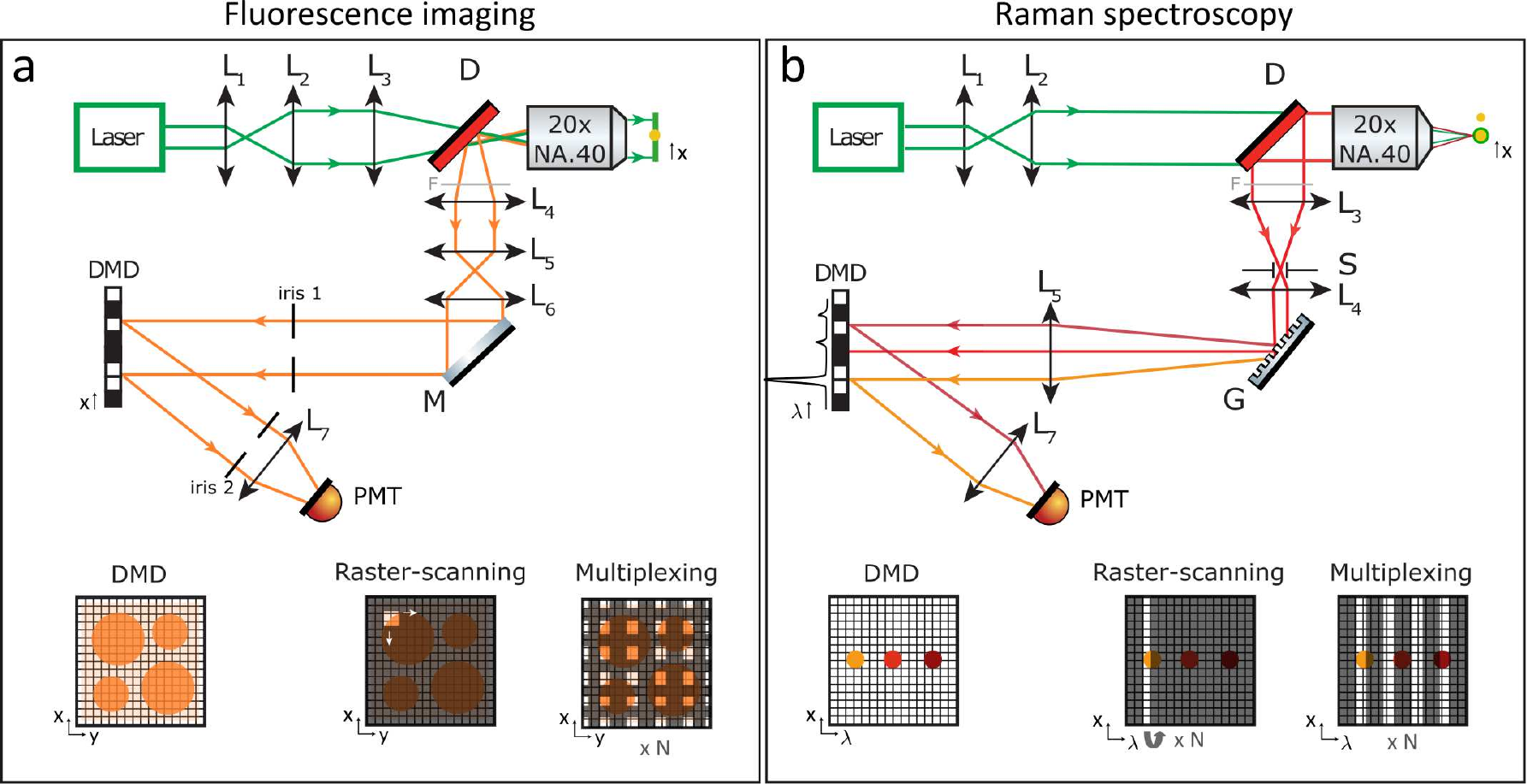} 
\caption{Schematic of the experimental setups.
(a) The fluorescence signal emitted by the  sample is imaged onto the DMD plane. $L_1$-$L_6$, convex lenses with focal lengths 50 mm, 150 mm, 150 mm, 150 mm, 180 mm, 50 mm, and 150 mm, respectively; $L_7$, combination of two lenses imaging the DMD into the PMT with $\times$3 demagnification (b) The wavelength components of the Raman signal emitted by the sample are dispersed onto the DMD.  Same components as in (a) except for $L_3$ 100 mm, $L_4$ 100 mm $L_5$ 150 mm. D, dichroic mirror; F, notch filter; M, mirror; PMT, photon-counting photomultiplier tube; S, confocal slit; G, amplitude grating}
\label{fig:setups}
\end{figure*}

For both RS and multiplexing, the setups layouts are similar, and both make use of a digital micromirror device (DMD). The DMD is a 2-D matrix of micromirrors, controlled to either direct the light to a detector or discard it to a beam dump. Since the DMD acts as a binary modulator, it is used to implement both raster-scanning and multiplexing: RS is performed by turning on each DMD pixel (or group of pixels) one-by-one, and multiplexing is performed by displaying each reshaped multiplexing-matrix row sequentially onto the DMD. 
In the dual detection scheme, each pattern and its complementary are displayed sequentially, which models the presence of a second detector (that would detect simultaneously the complementary measure).
For the 2-steps multiplexing scheme, the two patterns can be displayed sequentially, or obtained simply by displaying each reshaped row of a $\mathbf{M} \otimes \mathbf{M}$ matrix.

\subsubsection*{Widefield fluorescence imaging (Fig.~\ref{fig:setups}a)}
\label{sec: Widefield fluorescence imaging}

\noindent \textbf{Principle}:
The DMD plane contains the fluorescence signal emitted from the sample plane . Multiplexing the DMD pixels along (x,y) combines several spatial bins into the detector at each instant. Raster-scanning the DMD along (x,y) is formally equivalent to scanning the sample plane with a point-focus such as in (Fig.~\ref{fig:setups}a). We choose to perform raster-scanning onto the DMD instead of implementing it physically to make the SNR comparisons more reliable.  \\

\noindent \textbf{Experimental setup}:
On the illumination side, a continuous wave laser (532~nm Verdi, Coherent Inc) is focused onto the the back focal plane of a microscope objective (Olympus 20x, 0.4 NA) to create a widefield illumination in the sample plane. On the detection side, the fluorescent signal from the object, selected via a dichroic mirror and notch filter, is imaged with a x60-demagnification onto the DMD (V-7001, Vialux -$1024\times768$ mirrors). When the DMD pixels are in the 'ON' state, the signal impinging on these pixels is deflected into a photon-counting PMT (H7421-40, Hamamatsu).
The theoretical spatial resolution of the system is about 1~$\mu$m. 
The theoretical FOV is about 600~$\mu$m, but in practise we reduce it to 80~$\mu$m or 40~$\mu$m by using only a sub-part of the DMD area. In addition, an iris is placed before the DMD to limit the amount on light impinging on the DMD, and the associated spurious signal arising from pixels in the 'OFF' state. For the same reason, an iris is also placed right after the DMD, to only select the central diffraction order created by the blazed-grating structure of the device \cite{wavefronshapingNet_DMDdiffraction}. \\

\noindent \textbf{Sample}:
The sample is made of fluorescent particles of 15~$\mu$m (F36909 FocalCheck fluorescence microscope test slide 1 - invitrogen).  \\

\noindent \textbf{Excitation power and integration time}:
The experiments are carried at constant integration time and irradiance for raster-scanning and multiplexing. The maximum excitation intensity is chosen such as the maximum count rate lies within the linearity range of the detector ($\approx$ 10$^6$ counts/s). The laser power at the sample plane is about 50mW (irradiance $\approx$ 7.2 $\times$ 10$^{-9}$ W/$\mu$m$^2$). The exposure times are set to 10~ms per measurement.\\

\noindent \textbf{Number of measurements:}
The two fluorescent samples of Fig.~5 
are the same, but the FOV is more or less cropped to artificially render the sample more sparse. In the two cases, the spatial sampling is about 1.3~$\mu$m (DMD pixels binned 4-by-4 along x and y). 
\begin{itemize}
\item Sample 1: For one-step multiplexing and dual detection, we multiplex with a positive-Hadamard-matrix of size $N$ = 64 $\times$ 64 = 4096; for two-step multiplexing, we choose the matrix $\mathbf{S}\otimes \mathbf{S}$ with the closest dimensions with $N$ = 63 $\times$ 63 = 3969.
\item Sample 2: For one-step multiplexing and dual detection, we multiplex with a positive-Hadamard-matrix of size $N$ = 32 $\times$ 32 = 1024; for two-step multiplexing, $N$ = 31 $\times$ 31 = 961.
\end{itemize}

\noindent \textbf{Data processing}:
All the measurements are repeated 20 times in the exact same configurations to calculate statistical values. We choose to calculate the variance of the experimental estimation rather than the MSE. Indeed, we expect the differences between raster-scanning and multiplexing performances to be subtle, and with no access to the real ground truth, we do not want to favour one or the other with some potential experimental bias. \\
The multiplexing matrix is pre-compensated to take into account for the diamond-shape of the DMD (placed at 45 degrees) and avoid mismatch between the theoretical and physical multiplexing matrix \cite{Rodriguez2016}. \\
To calculate the experimental object mean $\bar{x}$ via:
\begin{equation}
\bar{x} = (\bar{x}_{RS} + \bar{x}_{H1} + \bar{x}_{H1b})/3 
\end{equation}
where $\bar{x}_{RS}$ is the object mean obtained by averaging all raster-scanning measurements, $\bar{x}_{H1}$ is the object mean obtained by averaging all one-step measurements, and $\bar{x}_{H1b}$ is the object mean obtained by averaging all dual-detection measurements. We discard two-steps multiplexing since it leads to the the highest error.  
The number of realisations being small, we apply a Gaussian filter with $\sigma$=1 on the variance section plots for clarity (but not on the variance images). The row variance plots are shown in Fig.~\ref{fig:Fluo_rawVar1D}. 

\begin{figure*} [hbt!]
\centering
\includegraphics[width=\linewidth]{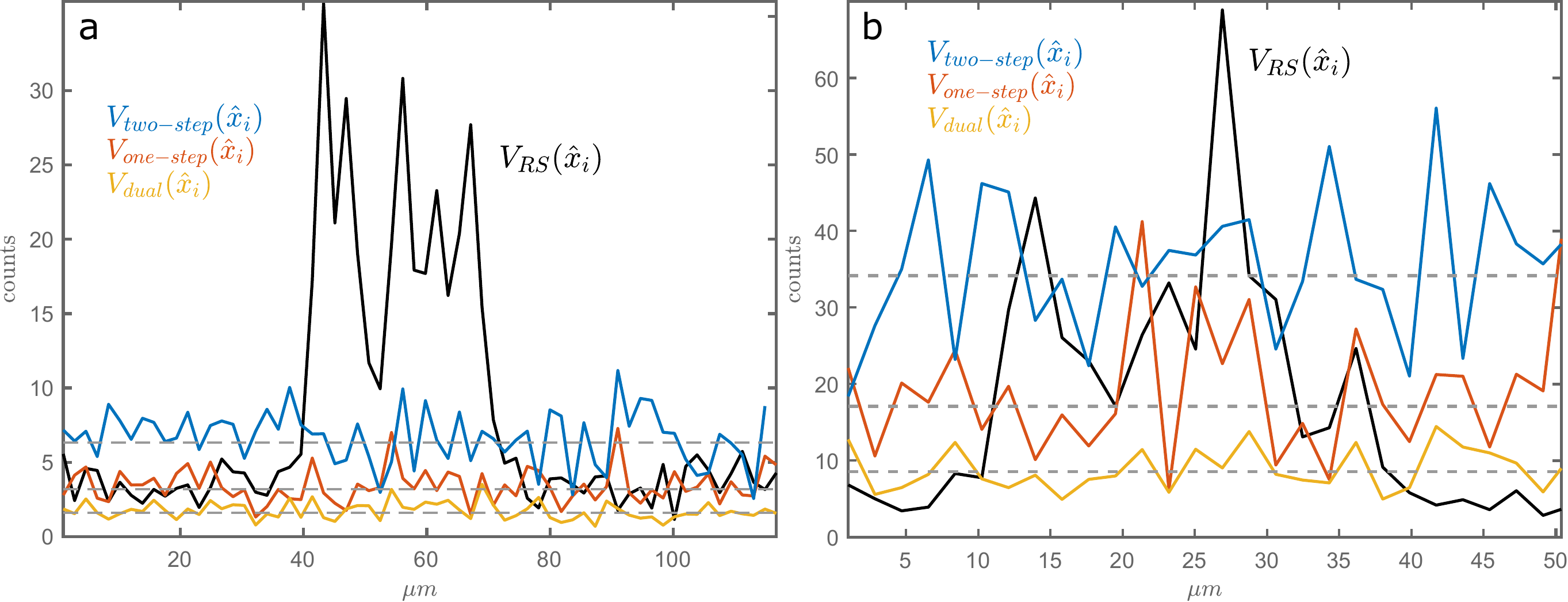} 
\caption{Raw variance sections (not Gaussian filtered), taken along the anti-diagonal of the variance images of Fig.~5 
c, f}
\label{fig:Fluo_rawVar1D}
\end{figure*}


\subsubsection*{Raman spectroscopy (Fig.~\ref{fig:setups}a)}

\textbf{Principle}:
The DMD $\lambda-$axis contains a Raman
spectrum. Raster-scanning the spectrum along $\lambda$ leads to a sequential measure of each wavelength bin. Instead, multiplexing sequentially measures combinations of several wavelengths. This is formally equivalent to comparing monochromators with either a moving exit slit or a coded-aperture spectrometer. \\

\noindent \textbf{Experimental setup}:
On the illumination side, a continuous wave laser (532~nm Verdi, Coherent Inc) is focused onto the sample plane, with a microscope objective (Olympus 20x, 0.4 NA). 
On the detection side, the Stokes Raman scattered light from the object is relayed onto a confocal slit. A combination of dichroic mirror and notch filter ensures only the Raman signal is retained. Next, it is dispersed with a blazed grating ($600~$mm$^{-1}$, Thorlabs), and the
spatially dispersed wavelength components are imaged onto the DMD (V-7001, Vialux -$1024\times768$ mirrors). The DMD $\lambda-$axis, in conjunction with the grating, acts as a programmable spectral filter. When the DMD pixels are in the 'ON' state, the signal impinging on these pixels is deflected into a photon-counting PMT (H7421-40, Hamamatsu), while the rest is sent into a beam dump. 
The spectral resolution of this system is about 40~ cm$^{-1}$; limited by the grating and the focal lengths lenses of the spectrometer. \\

\noindent \textbf{Sample}:
The sample is a liquid solvent - pure DMSO (Dimethyl Sulfoxide - 99.9\%, Sigma-Aldrich) - placed in a quartz spectroscopic cuvette. \\

\noindent \textbf{Excitation power and integration time}:
The experiments are carried at constant integration time and irradiance for raster-scanning and multiplexing. The maximum excitation intensity is chosen such as the maximum count rate lies within the linearity range of the detector ($\approx$ 10$^6$ counts/s). The laser power at the sample plane is about 1.2~mW (irradiance $\approx$ 3.3 $\times$ 10$^{-3}$ W/$\mu$m$^2$) and the exposure times are set to 5~ms per measurement. \\

\noindent \textbf{Number of measurements}:
The spectral resolution of the system allows us to bin the 1024 DMD pixels along $\lambda$-axis 8-by-8 with no resolution loss. This results in 128 effective pixels.
For one-step multiplexing, we multiplex with a S-matrix since in 1-D it is preferable over the positive Hadamard matrix (Table.~1 Main Text)
The identity matrix and S-matrix are of size 127$\times$127 ($N$ = 127). 
For dual detection, $N$ = 128.
For two-steps multiplexing, we choose the $\mathbf{S}\otimes \mathbf{S}$ with the closest dimensions, i.e. made of two S-matrices of size 11, thus $N$ = 121. 
Although this modality is not relevant in 1D, we model it to verify our theoretical results experimentally. \\

\noindent \textbf{Data processing}:
All the measurements are repeated $n$=1000 times in the exact same configurations to calculate the empirical means and variance. 
We choose to rather present results on the variance rather than on the MSE. Indeed, we expect the differences between raster-scanning and multiplexing performances to be subtle, and with no access to the real ground truth, we do not want to favour one or the other with some potential experimental bias.
We calculate the experimental object mean $\bar{x}$ in the same way than for fluorescent imaging:
\begin{equation}
\bar{x} = (\bar{x}_{RS} + \bar{x}_{S} + \bar{x}_{Sb})/3 \approx 10.6\text{ counts}
\end{equation}
(with $\bar{x}_{RS}= 11$, $\bar{x}_{S} =10.5$, $\bar{x}_{Sb} = 10.5$). 
On the variance plot (Fig.~5
j), the amplitude of the error bar at one standard deviation is (normal distribution approximation) \cite{Ruch2013}: 
\begin{equation}
\frac{2(n-1)}{n^2} (Var(\hat{x}_i))^2
\label{eqn:ErrorBar_Variance}
\end{equation}
where $Var(\hat{x}_i)$ the empirical variance obtained after estimation over all  $n=$1000 measurements. 


\subsubsection*{Photon-noise hypothesis}
\label{sec: Verification of the noise model}
The paper is based on the hypothesis that each measured number of photons $b$ is a random variable whose probability law is a Poisson distribution of mean $b_0$. To verify this hypothesis, the mean number of counted photoelectrons should be equal to its variance with $\langle b \rangle = \langle \delta b^2 \rangle = b_0$. \\
Experimentally, we count the detected photons through the spectroscopic system (Fig.~\ref{fig:setups}b) with a sample of DMSO, with all DMD pixels 'ON'. 
On Fig.~\ref{fig:Shot_noise}a, the laser power is fixed to 0.5~mW and the integration time is varied between 0.1~ms and 10~ms. 
Each measurement is repeated 2000 times, and we verify that variance approximately equals to the mean. In addition, the detector dark noise (Poisson distributed and signal independent) is measured 1000 times in complete darkness, for an exposure time of 1~s. Fig.~\ref{fig:Shot_noise}c shows the resulting normalized histogram, which can be fitted with a Poisson distribution of mean $\approx$ 9 (coherent with the PMT specifications). Thus, the dark count of our detector is about of 9 photoelectrons per second: since our integration times are of the order of 5-10~ms, this value is considered as negligible as compared to the typical count rates measured in the context of the present experiments. 

\subsubsection*{Noise model for the experimental data}
\label{sec: SI_Model for the measurements}

As seen on Fig.~5
a-f, the contrast of the DMD is not perfect: the pixels in the 'OFF' order contribute to some amount to the signal detected in the 'ON' order. This means that, even when all the DMD pixels are 'OFF', there is still a small portion $\alpha$ of the signal of the DMD plane (e.g. fluorescence, Raman) that contributes to the 'ON' order and therefore enters the detector. 
In our case, we estimate this relative contribution $\alpha$ to $\approx$ 0.065$\%$ (by measuring the ratio between the signal when the DMD is all 'ON' and all 'OFF', for different samples). 
Although this contribution seems insignificant, it may seriously impact the measurements. Indeed, if a 100 $\times$ 100 pixels object emits on average 10 photons per pixel, the object sum (i.e. signal all DMD 'ON') would account for $10^5$ counts, thus the 'OFF' order for $N \alpha \bar{x} =$ 65 counts, which may be more than the intensity of each pixel. 
The relative contribution $\alpha$ is independent of the object signal, but the absolute contribution of the 'OFF' order, equal to $N \alpha \bar{x}$ , depends on the object signal. \\

In Raman spectroscopy experiments (Fig.~5
g-j), the DMD 'OFF' order contributes to $ N \alpha \bar{x} \approx$ 0.9 counts ($N = 127$, $\bar{x}$ = 10.6 counts).  This contribution is negligible, as verified on Fig.~\ref{fig:dessin_checkModel}a: indeed the mean and variance for raster-scanning experiments are quasi-equal (difference of less than one count). 
Therefore, our Raman spectroscopy experiments can indeed be modelled by the simple initial model:  
\begin{equation*} 
\mathbf{b} \sim Poisson(\mathbf{A} \mathbf{x})
\end{equation*}

In the fluorescent imaging experiments of Fig.~5 
a-f, the contribution from the DMD OFF order cannot be considered as negligible (Fig.~\ref{fig:dessin_checkModel}b). For sample 1, $ N \alpha \bar{x} \approx$ 4.25 counts ($N = 4096$, $\bar{x}$ = 1.6 counts); for sample 2, $ N \alpha \bar{x} \approx$ 5.7 counts ($N = 1024$, $\bar{x}$ = 9 counts). \\
Therefore, the object is actually multiplexed by $\mathbf{A}$ plus an constant offset matrix $\alpha \mathbf{J}_N$
. For raster-scanning, this leads to $
\mathbf{b} \sim Poisson((1-\alpha)\mathbf{I}_N + \alpha \mathbf{J}_N ) \mathbf{x} )$, or to 
$\mathbf{b} \sim Poisson((\mathbf{I}_N + \alpha \mathbf{J}_N ) \mathbf{x} )$ since in our case, $\alpha << 1$. \\
For positive-Hadamard multiplexing, half of the pixels are 'ON' at each measurement, which leads to
$\mathbf{b} \sim Poisson(\mathbf{A}+ \alpha \mathbf{J}_N - \frac{\alpha}{2}\mathbf{J}_N) \mathbf{x} )$. 
Therefore, the general model can be written as: 
\begin{equation} 
\mathbf{b} \sim Poisson((\mathbf{A}+ \alpha \mathbf{J_N}) \mathbf{x} )
\label{eqn:ModelwithCteMatrix}
\end{equation}
with $\alpha$ for raster-scanning, $0.5 \alpha$ for one-step multiplexing and $0.75 \alpha$ for two-steps multiplexing with S-matrices. The LS-estimation is thus performed by inverting the matrix $\mathbf{A}+ \alpha \mathbf{J_N}$.
Note the experiment could as well be modelled as $\mathbf{b} \sim Poisson(\mathbf{A}  \mathbf{x} + \bm{\eta})$, with $\bm{\eta}=\alpha \mathbf{J}_N \mathbf{x} = N \alpha \bar{x}$. Yet, this implies a calibration step for each new sample, in order to estimate $\bm{\eta}$.  More details are given in Supplementary methods and in \cite{Scotte2022}. \\
We emphasize that it is crucial to take this imperfection into account into the model to compare raster-scanning and multiplexing (Fig.~\ref{fig:dessin_nobckmodel}). Otherwise, the raster-scanning results would be biased by a factor $ N \alpha \bar{x}$.

\begin{figure*} [h]
\centering
\includegraphics[scale=0.65]{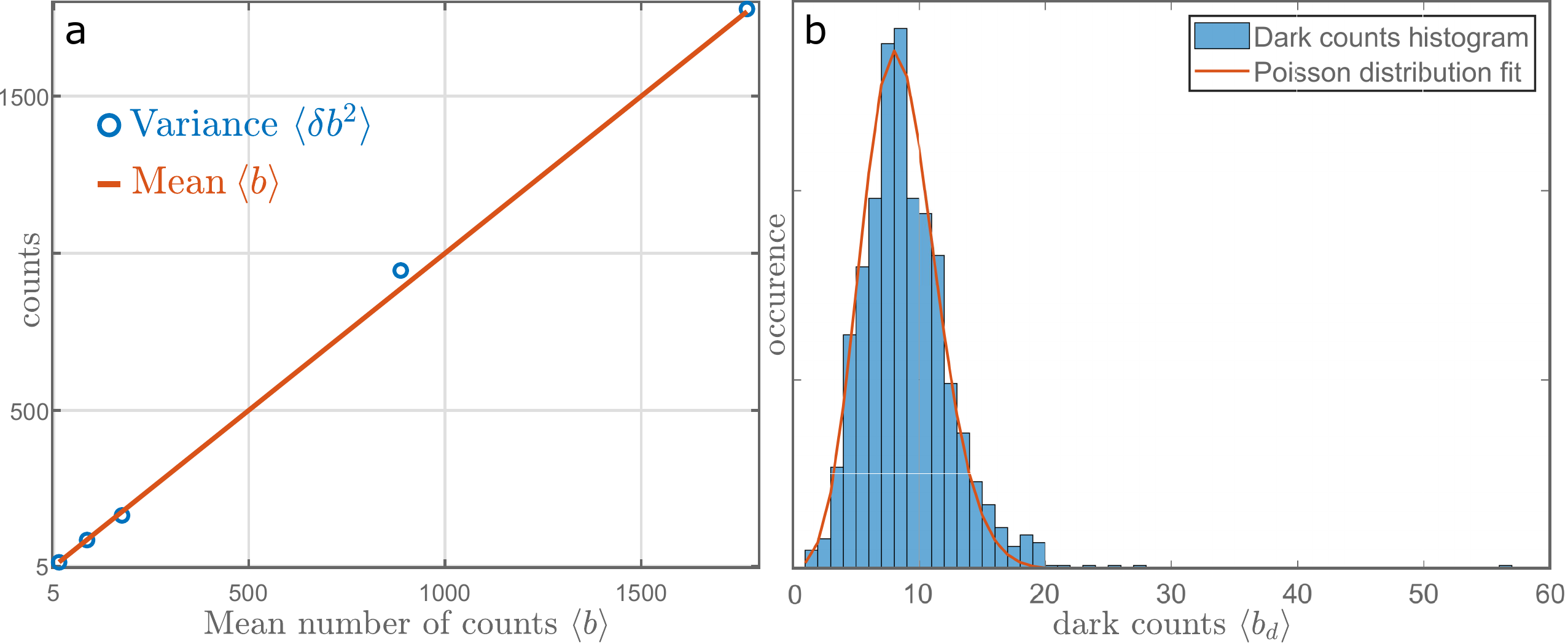} 
\caption{(a) Experimental mean and variance of the counted photons, at fixed laser power; (b) normalized histogram of measured detector dark-counts}
\label{fig:Shot_noise}
\end{figure*}

\begin{figure*} [h!]
\centering
\includegraphics[width=\linewidth]{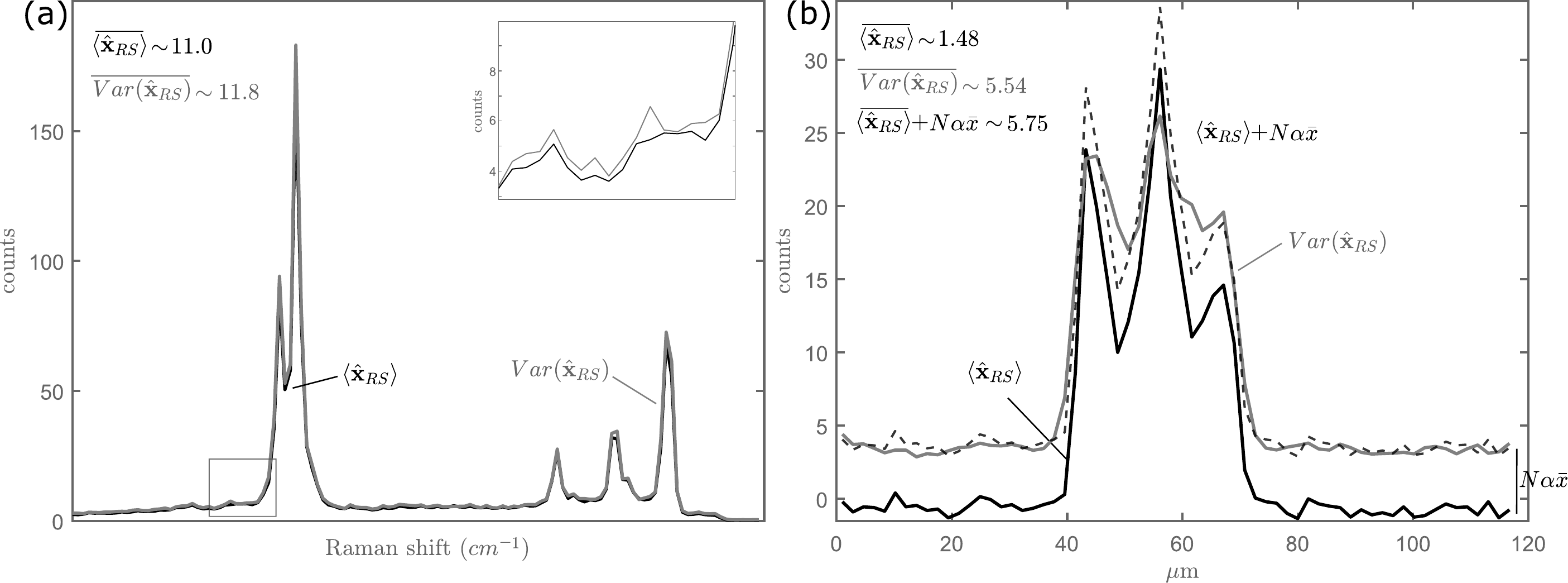} 
\caption{Verification of the proposed models for one-step multiplexing with the samples of Fig.~5.
(a) The mean and variance of a DMSO Raman spectrum differ from less than one count (b) Section (along the anti-diagonal) of the mean and smoothed variance of the fluorescent object (sample 1). The  variance is equal to the estimation mean plus an offset equal to $N \alpha \bar{x}$.  Note that the negative values of (b) are due to the inversion of the matrix $\mathbf{A}+ \alpha \mathbf{J}_N$.}
\label{fig:dessin_checkModel}
\end{figure*}

\begin{figure*} [h!]
\centering
\includegraphics[width=\linewidth]{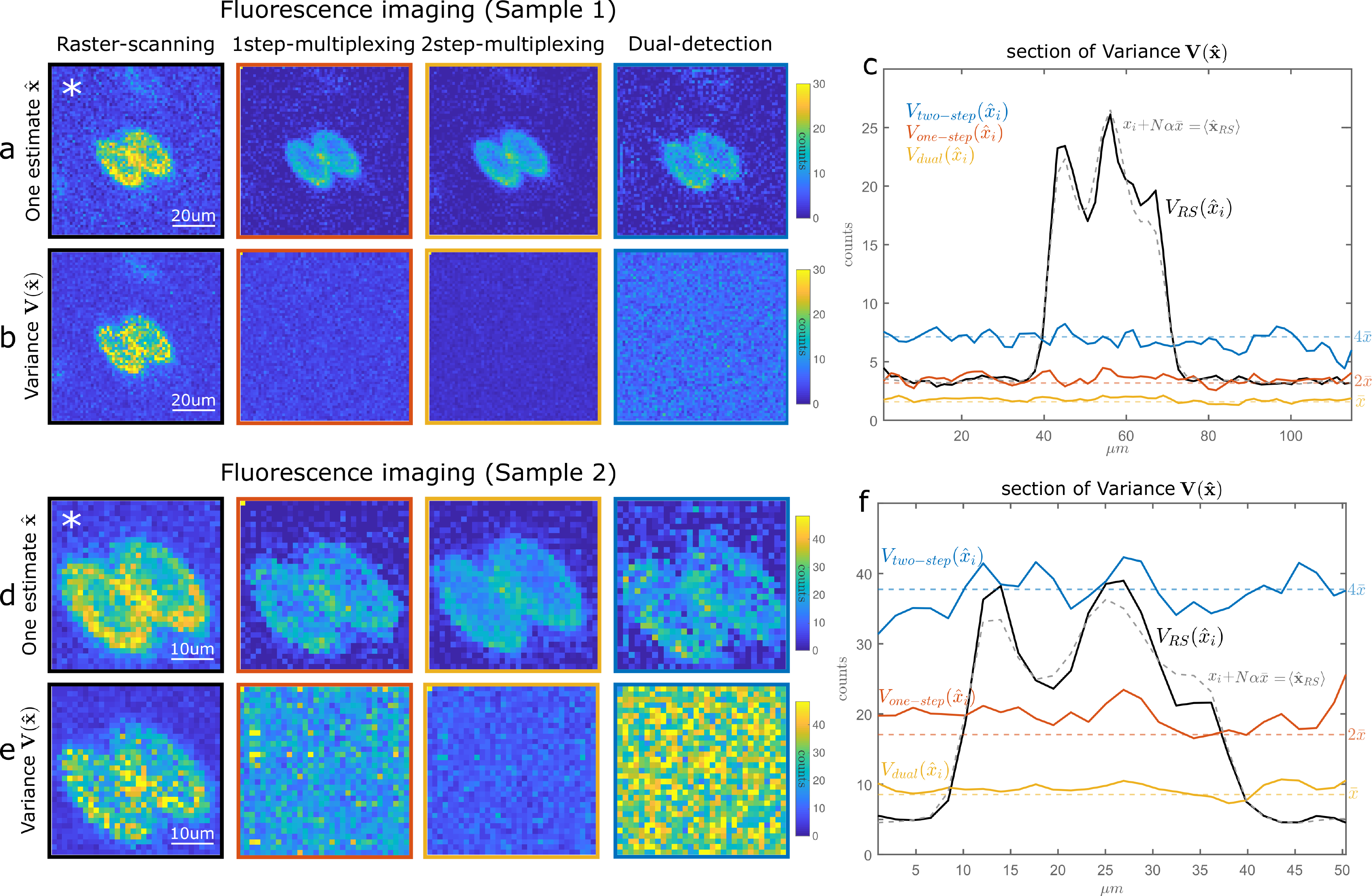} 
\caption{Same results as Fig.~5
a-f, but without correcting the model as in equation~\eqref{eqn:ModelwithCteMatrix}. Since here, $\alpha \bar{x}$ is sufficiently small, only the raster-scan estimation (*) is changed: the estimation is biased by an offset equal to $N \alpha \bar{x}$. }
\label{fig:dessin_nobckmodel}
\end{figure*}

\cleardoublepage

\section*{Supplementary figures}
\begin{figure*} [h!]
\centering 
\includegraphics[width=\linewidth]{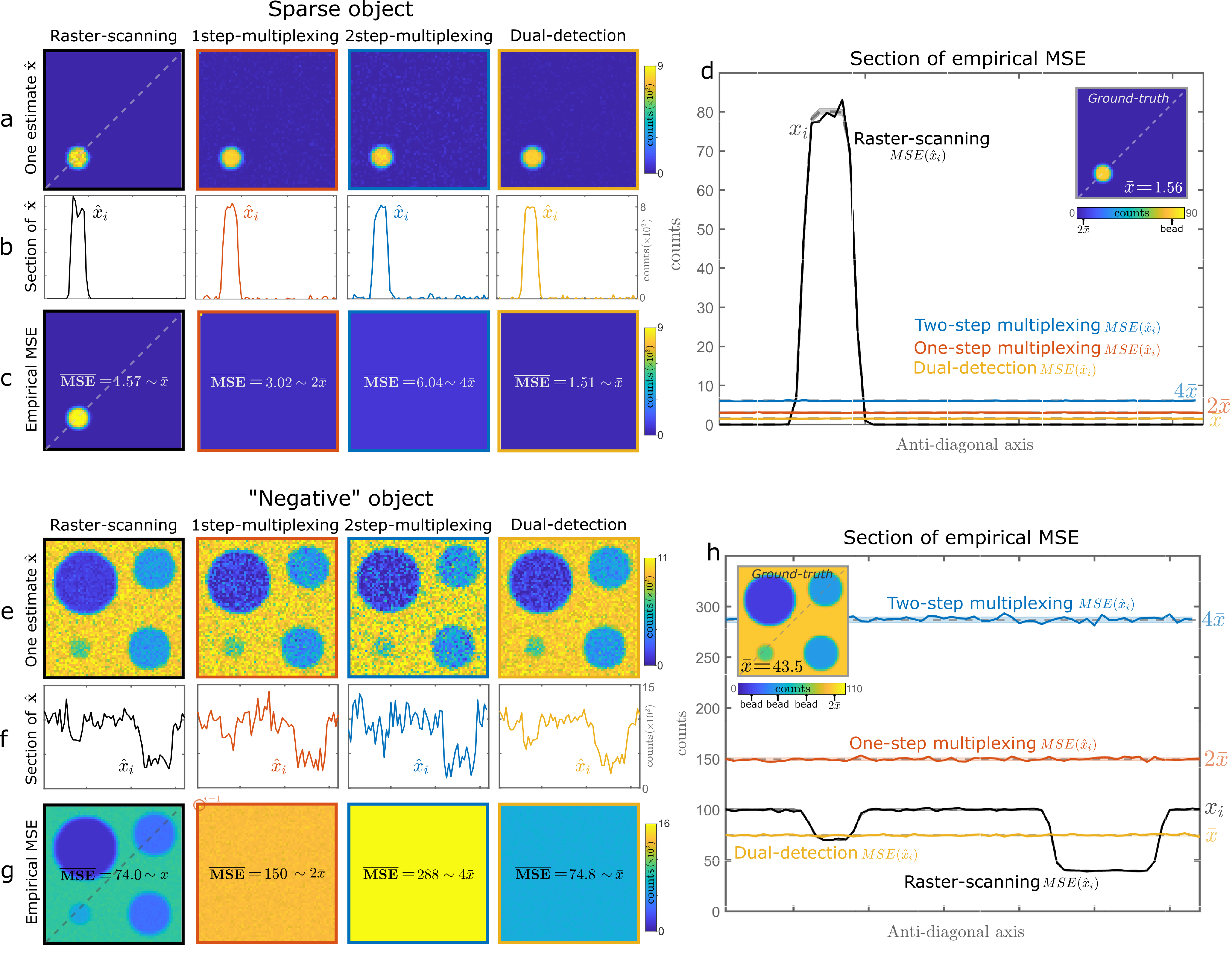} 
\caption {MSE for positive-Hadamard multiplexing, for a sparse and "negative" object. One-step multiplexing is performed with $\mathbf{H_1}$, two-step with $\mathbf{S} \otimes \mathbf{S}$, and dual detection with $\mathbf{H_1} - \mathbf{H_2}$ ($\mathbf{H_2}$ is the complementary of $\mathbf{H_1}$).
\textbf{a, e}, Example of one estimate $\hat{\mathbf{x}}$ obtained after one realisation of the data (one simulated measurement and LS-estimation). For visualisation purposes, only positive estimated values are shown.
\textbf{b, f}, Section of $\hat{\mathbf{x}}$ along the anti-diagonal. 
\textbf{c, g}, Empirical MSE obtained with $n=20,000$ noise realisations. $\overline{MSE}$ is its average over all pixels (except the first pixel for one-step multiplexing, see Table.1).
\textbf{d, h}, Section of $\mathbf{MSE}(\hat{\mathbf{x}})$ along the anti-diagonal. The dashed-lines represent theoretical MSE values, and the faint strip the error-bar. Inset: ground-truth object $\mathbf{x}$. }
\label{fig:SI_dessin_OtherObjects}
\end{figure*}

\begin{figure*} [h]
\centering
\includegraphics[width=\linewidth]{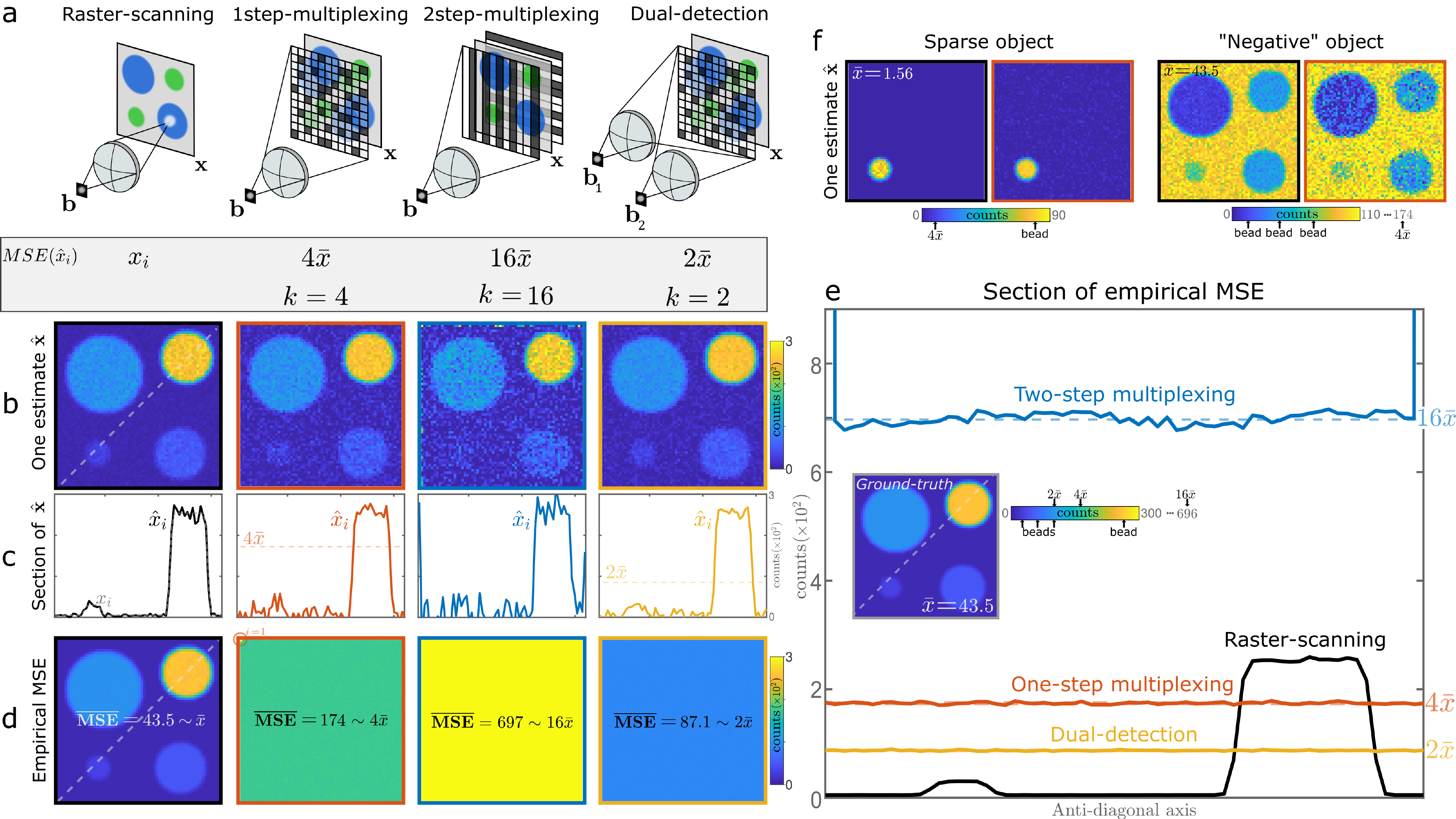} 
\caption{MSE for positive-cosine multiplexing with the matrix $\mathbf{C_1}$. 
\textbf{a}, Schematic representation of a 2-D imaging system in raster-scanning mode and in three multiplexing modes, with the associated theoretical MSE per object pixel $i$ and constant $k$. 
\textbf{b}, Example of one estimate $\hat{\mathbf{x}}$ obtained after one realisation of the data (one simulated measurement and LS-estimation). For visualisation purposes, only positive estimated values are shown.
\textbf{c}, Section of $\hat{\mathbf{x}}$ along the anti-diagonal. The dashed-lines represent theoretical MSE values. 
\textbf{d}, Empirical MSE obtained with $n=20,000$ noise realisations. $\overline{MSE}$ is its average over all pixels (except the first pixel for one-step multiplexing, and the first row and column of pixels for two-step multiplexing, see Table. 2 
\textbf{e}, Section of $\mathbf{MSE}(\hat{\mathbf{x}})$ along the anti-diagonal. The dashed-lines represent theoretical MSE values. 
inset: ground-truth object $\mathbf{x}$ of intensity average $\bar{x}$. 
\textbf{f}, Example of one estimate for a sparse object and a "negative object" (object with structures of interest dimmer than a bright background), for raster-scanning and one-step multiplexing.
}
\label{fig:SimuResultsDCT}
\end{figure*}

\begin{figure*} [hbt!]
\centering
\includegraphics[width=\linewidth]{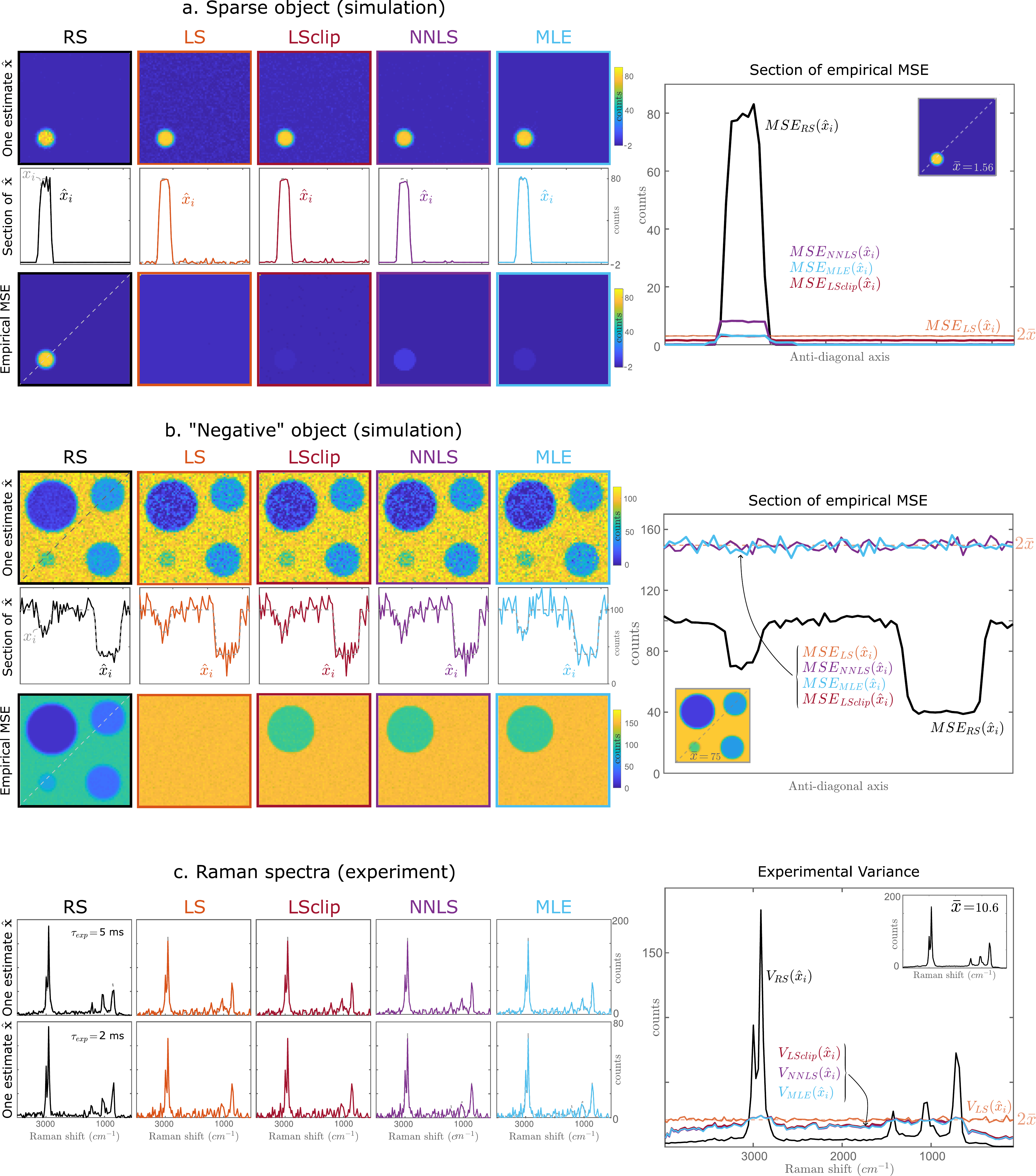} 
\caption {Effect of the estimators as compared on the MSE, for positive-Hadamard multiplexing and one-step multiplexing. LS: least-square, LS-clip: least-square with positive threshold, NNLS: non-negative least-square, MLE: maximum likelihood estimator. \textbf{a, } Effect on a sparse object. LS-clip discards the negative values and improves the MSE. NNLS and MLE further reduce the MSE on the background, at the expense of an increase on the peak for NNLS (Supplementary methods). $n = 5,000$ noise realisations. \textbf{b, } Effect on a "negative" object. This object is bright therefore the positivity-constraint hardly applies: other estimators give approximately the same MSE than LS. $n = 5,000$ noise realisations. \textbf{c, } Effect on the experimental Raman spectra. The spectrum is not sparse (presence of a positive background). LS-clip allows to discard negative estimated values, but the other estimators do not bring an additional improvement. $n = 1,000$ noise realisations.}
\label{fig:SI_Effect_estimators_Allobjects}
\end{figure*}

\begin{figure*} [hbt!]
\centering
\includegraphics[scale=0.7]{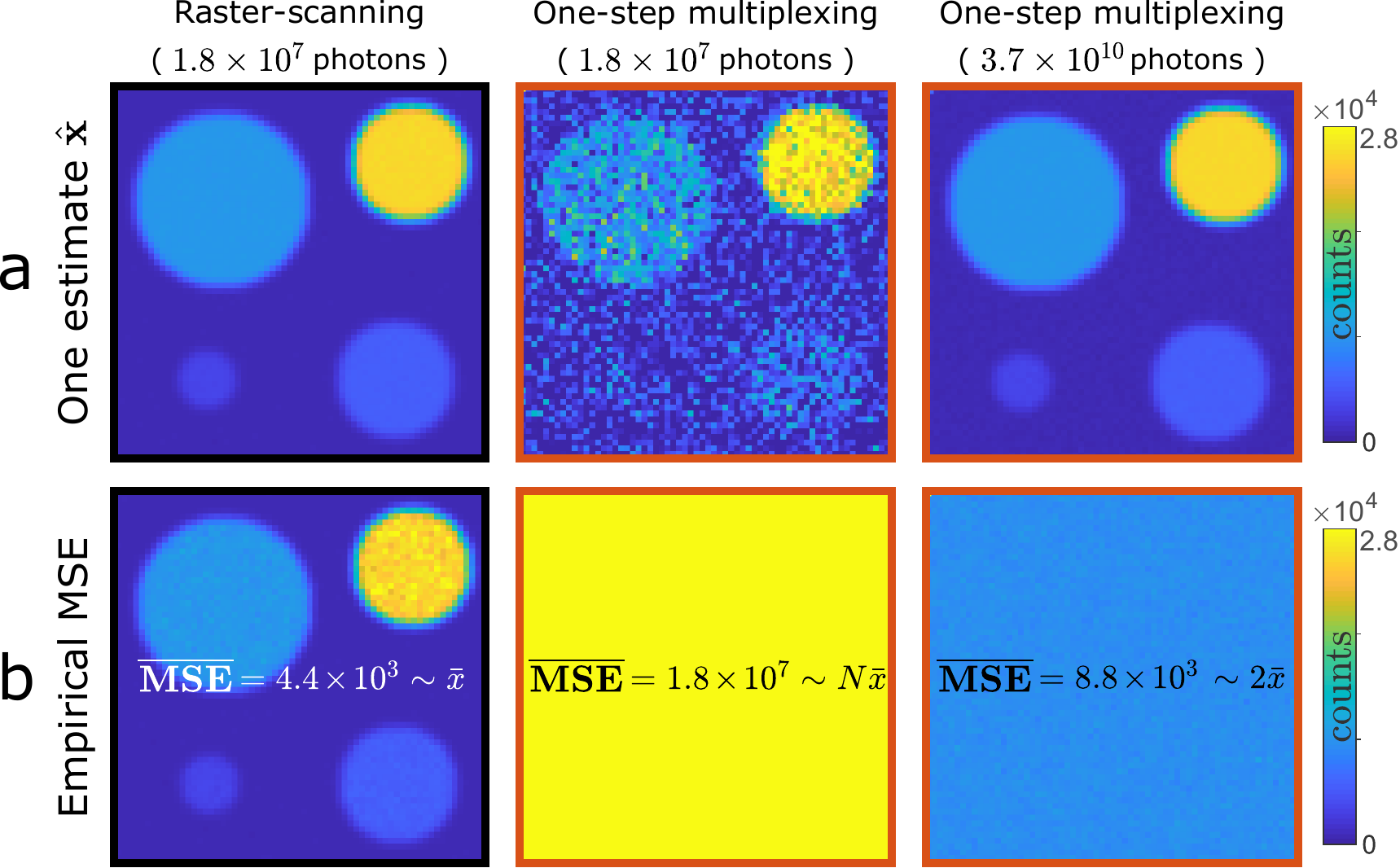} 
\caption {Comparison of raster-scanning and positive-Hadamard multiplexing at constant number of photons ($1.8\times 10^7$ collected photons in total in both cases), and at constant irradiance and exposure time (2000 more photons detected in  multiplexing).
\textbf{a, } At constant number of photons, the SNR of positive-Hadamard multiplexing is significantly degraded \textbf{b, } by a factor proportional to the number of pixels $N$, (equation 11 Main text)
$N=4096$, $\bar{x}=4.4\times 10^3$.}
\label{fig:CtNbPhotons}
\end{figure*}


\cleardoublepage
%


\bigskip
\bibliography{Ref_MainText}
\bibliographystyle{unsrt}

\end{document}